\pdfoutput=1
\documentclass{optica-article}

\journal{ome}

\articletype{Research Article}

\usepackage{graphicx}
\usepackage{bm}
\usepackage{float}
\usepackage{caption}
\usepackage{subcaption}
\usepackage{amsmath}
\usepackage[latin1]{inputenc}
\usepackage{mathtools}
\usepackage{gensymb}
\usepackage{upgreek}
\usepackage{orcidlink}
\usepackage{xcolor}

\usepackage{lineno}

\begin{document}
\title{Lasing at a Stationary Inflection Point}

\author{
A.~Herrero-Parareda~\orcidlink{https://orcid.org/0000-0002-8501-5775}\authormark{1}, 
N.~Furman~\orcidlink{https://orcid.org/0000-0001-7896-2929}\authormark{1}, 
T.~Mealy~\orcidlink{https://orcid.org/0000-0001-7071-9705}\authormark{1}, 
R.~Gibson~\orcidlink{https://orcid.org/0000-0002-2567-6707}\authormark{2}, 
R.~Bedford~\orcidlink{https://orcid.org/0000-0002-0457-0081}\authormark{3}, 
I.~Vitebskiy~\orcidlink{https://orcid.org/0000-0001-8375-2088}\authormark{2}, and 
F.~Capolino~\orcidlink{https://orcid.org/0000-0003-0758-6182}\authormark{1,*}
}

\address{\authormark{1}Department of Electrical Engineering and Computer Science, University of California, Irvine, CA 92617, USA\\
\authormark{2}Air Force Research Laboratory, Sensors Directorate, Wright-Patterson Air Force Base, Ohio 45433, USA \\
\authormark{3}Air Force Research Laboratory, Materials and Manufacturing Directorate, Wright-Patterson Air Force Base, Ohio 45433, USA}
\email{\authormark{*}f.capolino@uci.edu} 



\begin{abstract}
The concept of lasers based on the frozen mode regime in active periodic optical waveguides with a 3rd-order exceptional point of degeneracy (EPD) is advanced. The frozen mode regime in a lossless and gainless waveguide is associated with a stationary inflection point (SIP) in the Bloch dispersion relation, where three Bloch eigenmodes coalesce forming the frozen mode. As a practical example, we consider an asymmetric serpentine optical waveguide (ASOW). An ASOW operating near the SIP frequency displays a large group delay of a non-resonant nature that scales as the cube of the waveguide length, leading to a strong gain enhancement when active material is included. Therefore, a laser operating in the close vicinity of an SIP has a gain threshold that scales as a negative cube of the waveguide length. We determine that this scaling law is maintained in the presence of small distributed losses, such as radiation associated with waveguide bends and roughness. In addition, we show that although gain causes a distortion in the modes coalescing at the SIP, the properties of the frozen mode are relatively resistant to such small perturbations  and we still observe a large degree of exceptional degeneracy for gain values that bring the system above threshold. Finally, our study also reveals that lasing near an SIP is favored over lasing near a photonic band edge located in close proximity to the SIP. In particular, we observe that an SIP-induced lasing in an ASOW displays lower gain threshold compared to lasing near the photonic regular band edge (RBE), even though the SIP resonance has a lower quality factor than the RBE resonance. 
\end{abstract}

\section{Introduction}
\begin{figure}[t]
 \begin{subfigure}[t]{.64\textwidth}
  \centering
  \includegraphics[width=\linewidth]{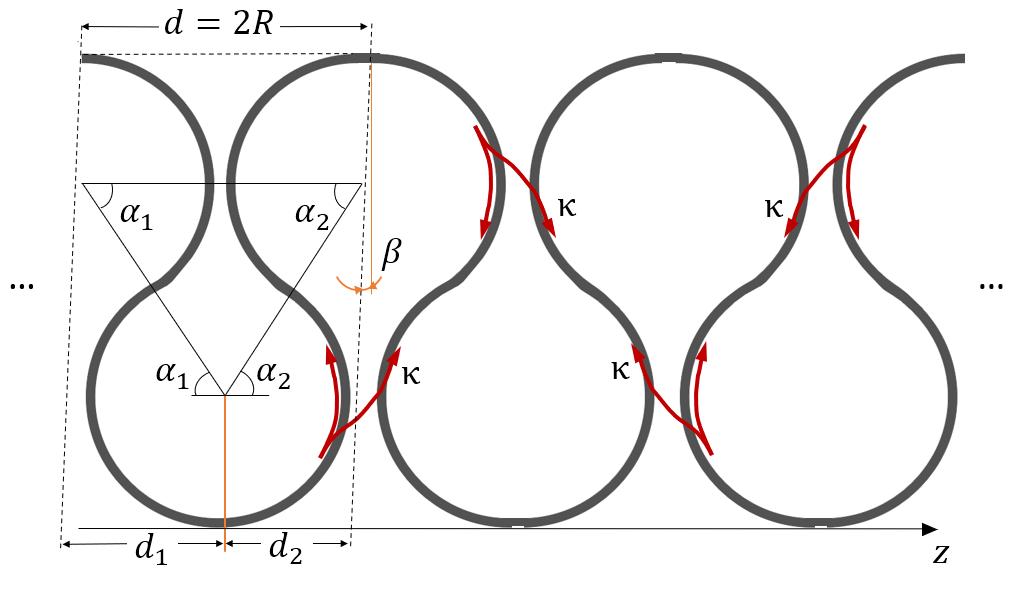}
  \caption{}
 \end{subfigure}
 \begin{subfigure}[t]{.38\textwidth}
  \centering
  \raisebox{-0.02\height}{\includegraphics[width=\linewidth]{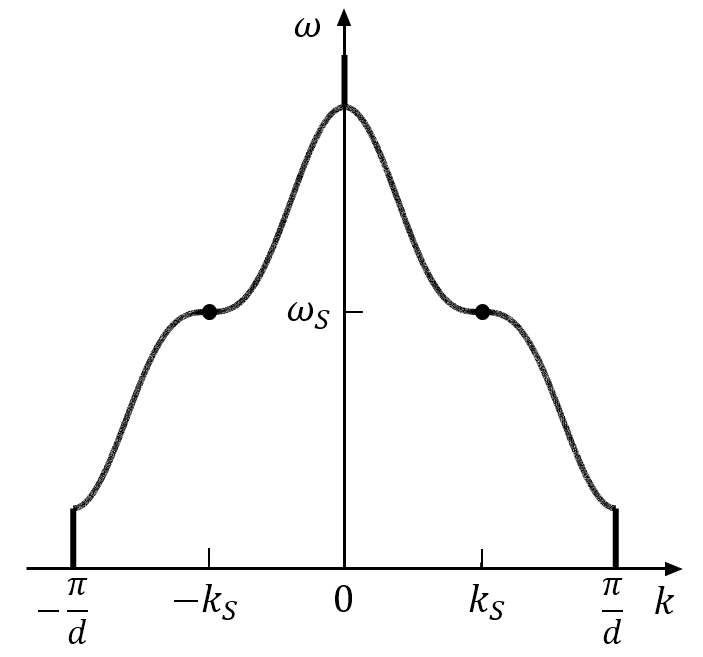}}
  \caption{}
 \end{subfigure}
 \caption{(a) SIP laser made of a periodic ASOW; gain is assumed to be distributed uniformly along the waveguide. The SOI waveguide follows a serpentine path. A unit cell of length $d=2R$ is defined within the two oblique dashed lines as in \cite{parareda_frozen_2021}. There are bidirectional coupling points between adjacent loops (denoted by red arrows). (b) Dispersion diagram of the propagating modes, with SIPs at $\omega_S$.}
\label{fig:Fig1Waveguide}
\end{figure}

An exceptional point of degeneracy (EPD) is a point in a parameter space associated with the coalescing of the eigenvalues and the eigenvectors of a system, see for example Ref.~\cite{lancaster_eigenvalues_1964, kato_perturbation_1966, seyranian_sensitivity_1993, heiss_exceptional_2004, ruter_observation_2010, heiss_physics_2012}, or~\cite{figotin_slow_2006} where Figotin and Vitebskiy refer to EPDs as stationary points, without explicitly using the term EPD but providing detailed math and physics aspects. It has been shown that EPDs in photonic systems exist when waveguide modes are coupled in the presence of gain and loss, as in PT-symmetric systems \cite{ruter_observation_2010, ramezani_unidirectional_2010, schindler_experimental_2011, gear_parity_2015, othman_theory_2017, abdelshafy_exceptional_2019}. However, systems do not need to have loss and gain to exhibit an EPD. In this paper, indeed, we focus on the stationary inflection point (SIP) formation in periodic lossless and gainless waveguides \cite{figotin_electromagnetic_2003, stephanson_frozen_2008, mumcu_lumped_2010, apaydin_experimental_2012, gutman_frozen_2012, ramezani_unidirectional_2014, nada_theory_2017, volkov_unidirectional_2021, paul_frozen_2021, tuxbury_nonresonant_2022}. The SIP is an EPD of order three \cite{nada_theory_2017,nada_frozen_2021}. In lossless and gainless waveguides, a second, third, and a fourth-order exceptional degeneracy of Floquet-Bloch eigenmodes are associated with a regular band edge (RBE), an SIP \cite{figotin_oblique_2003, nada_theory_2017}, and a degenerate band edge (DBE) \cite{figotin_frozen_2006} in the frequency-wavenumber dispersion relation of the waveguide modes, respectively. In the vicinity of an EPD of order $m$ at $\left(k_m, \omega_m \right)$, the frequency-wavenumber dispersion relation is approximated as \cite{gutman_slow_2012, nada_theory_2017}

\begin{equation}
  (\omega-\omega_m) \propto (k - k_m)^m,
  \label{eq:DispRelmEPD}
\end{equation}

which shows that the higher the order of the EPD, the flatter the dispersion diagram in its vicinity. For example, an SIP ($m=3$) has a flatter dispersion diagram than an RBE ($m=2$). The DBE has a flatness given by $m=4$ and it is a degenerate version of the RBE. In contrast to the SIP, the RBE and the DBE have a bandgap on one side of the RBE or DBE frequency, therefore the SIP exhibits physical properties that differ from those of the RBE and DBE. The flatness at the SIP frequency $\omega_S$ is associated with the inflection point developed in the dispersion relation of the propagating component of the frozen mode \cite{figotin_electromagnetic_2013}, as is illustrated in Fig.~\ref{fig:Fig1Waveguide}(b). The flatness at the RBE frequency, however, is due to the coalescence of two counter-propagating modes. The common slow-wave resonances near an RBE are fundamentally different than the resonances that couple to the SIP-associated frozen mode because the SIP is not sided by a bandgap \cite{figotin_slow_2011} and the frozen mode is composed also by evanescent waves. Moreover, the SIP condition is not particularly sensitive to the size and shape of the underlying photonic structure \cite{li_frozen_2017}. The frozen mode regime exhibits exceptional properties including low group velocity, greatly enhanced quality factor and group delay characterized by a strong scaling with the waveguide length \cite{figotin_slow_2011}.

An enhanced group delay allows for large one-pass amplification compared to conventional lasers of the same cavity length \cite{othman_giant_2016}. This boost has been observed theoretically in lasing systems operating near an RBE \cite{dowling_photonic_1994} and a DBE \cite{othman_giant_2016, veysi_degenerate_2018}. The concept of unidirectional lasing near SIP in a periodic non-reciprocal multilayered structure was discussed in \cite{ramezani_unidirectional_2014}. A fundamental problem with the SIP realization in multilayered structures is that it essentially requires nonreciprocity, which at optical wavelengths is too small to achieve a well-defined SIP in the Bloch dispersion relation. In other words, the model considered in \cite{ramezani_unidirectional_2014} can be practically relevant only at microwave frequencies, where the nonreciprocal effects can be strong enough. By contrast, SIP realization in three-way periodic optical waveguides does not require the use of magneto-optical materials and can be achieved in perfectly reciprocal systems at optical wavelengths. We advance the theory of SIP lasing in a silicon-on-insulator~(SOI) reciprocal waveguide, namely in the periodic asymmetric serpentine optical waveguide~(ASOW). The ASOW is examined using coupled-mode theory and the transfer matrix method as was done for the waveguides studied in \cite{scheuer_serpentine_2011, nada_theory_2017}, and therefore our analysis is restricted to the linear regime. In the future, we aim to expand on the findings presented in this article by conducting time-domain simulations to provide an insight into mode selectivity and the impact of nonlinear effects in SIP lasers. The principal result of this paper is that resonances near the SIP frequency experience giant power gain when an active material \cite{siegman_lasers_1986} is integrated into the periodic structure. We establish that the lasing threshold corresponding to these resonances scales as $N^{-3}$, where $N$ denotes the number of unit cells in the finite-length waveguide. The presence of distributed losses, such as those resulting from the roughness or radiation of waveguide bends, naturally leads to an increase in the lasing threshold. For small distributed losses, however, the inverse cubic scaling of the lasing threshold with waveguide length remains intact. In fact, we demonstrate that even though small values of gain and loss distort the dispersion diagram of the modes near the SIP, due to the exceptional sensitivity of the mode wavenumbers at an EPD \cite{figotin_slow_2006}, the distorted modes retain similar characteristics as those exhibited by the frozen mode. The SIP condition is shown to be superior to the RBE condition for enhancing the power gain in cavities with the same gain medium and length. We observe that the SIP resonance induces a lower lasing threshold than the RBE resonance, even if the SIP resonance has a lower quality factor.

The paper is organized as follows: in Section~\ref{ch:LosslessGainlessSIPinASOW} the ASOW is defined as in \cite{parareda_frozen_2021} and we show an example of an SIP in ASOW. In Section~\ref{ch:LasingTheory} a gain medium is added to the ASOW. The effect of bending losses and gain-induced mode distortion on laser performance is analyzed in Section~\ref{ch:PracticalConsiderations}. In Section~\ref{ch:MultiEPD} we study potential lasing in the vicinity of RBEs close to the SIP and introduce design considerations to prevent it. The results are summarized in Section~\ref{ch:Conclusions}.

\section{SIP in ASOW}
\label{ch:LosslessGainlessSIPinASOW}
\begin{figure}
\centering
\includegraphics[width = 0.5\textwidth]{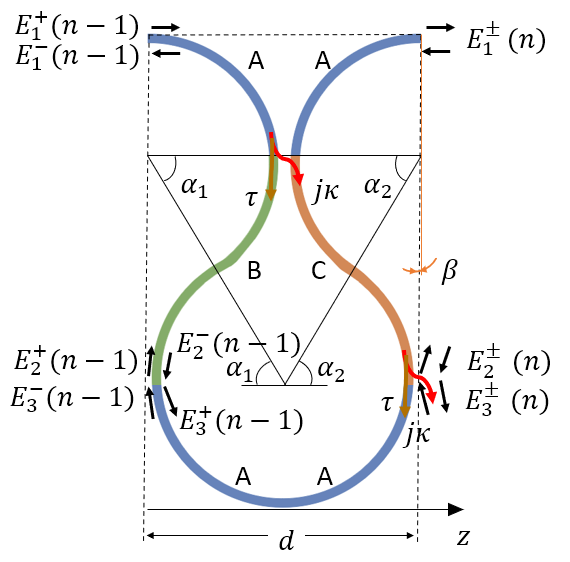}
\caption{The $n$-th unit cell of the ASOW is divided in four Sections A, a quarter of a circle long; Sections B and C, the arcs connecting the top and bottom loops at the left and right sides of the unit cell, respectively. The angles $\alpha_1$ and $\alpha_2$ determine the lengths of Sections B and C (respectively), and their difference $\beta$ determines the asymmetry of the unit cell. Besides the coupling to adjacent unit cells via Port 1, at each left and right sides of the unit cell, there are also proximity coupling points through Ports 2 and 3, at each side of the unit cell.}
\label{fig:Fig2UnitCell}
\end{figure}

The ASOW, introduced in \cite{parareda_frozen_2021} as a modification of the symmetric serpentine optical waveguide in \cite{scheuer_serpentine_2011}, is a SOI periodic asymmetric serpentine optical waveguide depicted in Fig.~\ref{fig:Fig1Waveguide}(a). It is composed of small rings of radius $R$ connected at angles $\alpha_1$ and $\alpha_2$, which are defined at the center of the rings and with respect to the horizontal axis. These rings are also side-coupled to the adjacent rings. Here, the evanescent coupling between adjacent rings is considered point-like, bidirectional, and lossless, as traditionally assumed \cite{vahala_optical_2004, yariv_photonics_2007}. The lossless condition ensures $\kappa^2 + \tau^2 = 1$, where $\kappa$ and $\tau$ are the coupling and transmission coefficients, respectively. The unit cell of the ASOW is defined between two oblique lines that descend from the top of two adjacent rings at an angle $\beta = \alpha_1 - \alpha_2$. The unit cell has a length of $d = 2R$, and it is portrayed in Fig.~\ref{fig:Fig2UnitCell}. It is convenient to split the unit cell into three different sections as was done in \cite{parareda_frozen_2021}: Section A, of which there are four per unit cell, describes a quarter of a circle and adds a phase of $\phi_{A} = k_0 n_w \pi R / 2$ to the waves traveling through it. The term $k_0 = \omega / c$ is the wavenumber in vacuum where $\omega$ is the angular frequency, $c$ is the speed of light in vacuum, and $n_w$ is the mode effective refractive index. Sections B and C describe the arcs connecting the top loops with the bottom loop, at the left and right sides of the unit cell, at angles $\alpha_1$ and $\alpha_2$ from the horizontal axis, respectively. They describe the asymmetry of the ASOW through the angle difference $\beta= \alpha_1 - \alpha_2$, and have an associated phase of $\phi_{B} = k_0 n_w 2 \alpha_1 R$ and $\phi_{C} = k_0 n_w 2 \alpha_2 R$. 
For an SIP to form, three eigenmodes must collapse on each other (in terms of wavenumber and polarization states, i.e., eigenvalues and eigenvectors), which requires them to belong to the same irreducible wavevector symmetry group. A way to do that is by setting $\beta\neq0$ \cite{parareda_frozen_2021}, which breaks the glide symmetry of the ASOW, though the SIP has been found also in glide symmetric waveguides \cite{nada_frozen_2021}. 

The electromagnetic guided fields are modeled in terms of the forward and backward waves \cite{parareda_frozen_2021}. Figure~\ref{fig:Fig2UnitCell} shows the three ports at both the $n-1$ and $n$ sides of the unit cell, and the two waves defined at the right side of each port. The state vector $\boldsymbol{\text{$\psi$}}(n)$ with all the six electric field wave amplitudes is defined as

\begin{equation}
  \boldsymbol{\text{$\psi$}}(n)=\left(\begin{array}{cccccc}
  E_{1}^{+}, \hspace{0.2cm}  E_{1}^{-}, \hspace{0.2cm} E_{2}^{+}, \hspace{0.2cm} E_{2}^{-}, \hspace{0.2cm} E_{3}^{+}, \hspace{0.2cm} E_{3}^{-}\end{array}\right)^T,
  \label{eq:StateVector}
\end{equation}

\begin{figure}[t]
 \begin{subfigure}[t]{.49\textwidth}
  \centering
  \includegraphics[width=\linewidth]{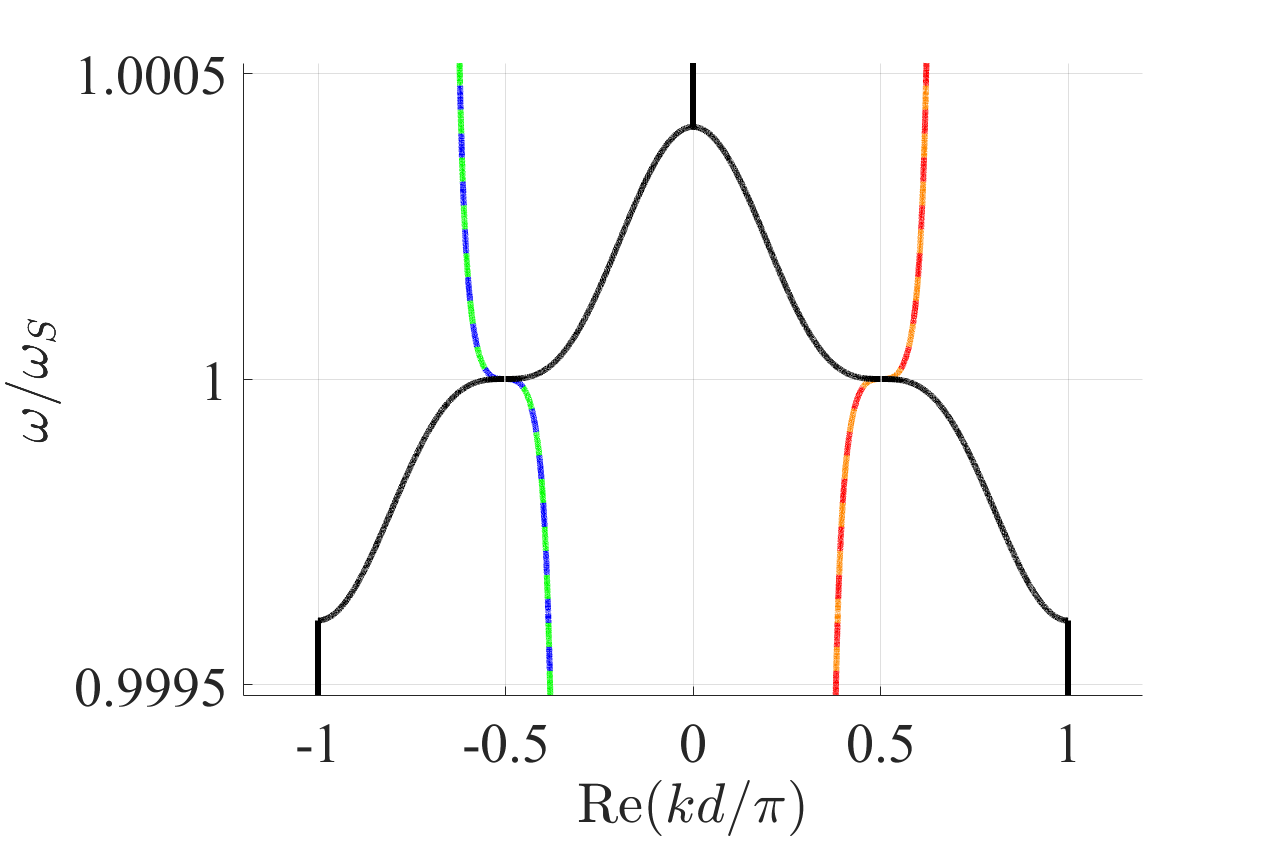}
  \caption{}
 \end{subfigure}
 \hfill
 \begin{subfigure}[t]{.49\textwidth}
  \centering
  \includegraphics[width=\linewidth]{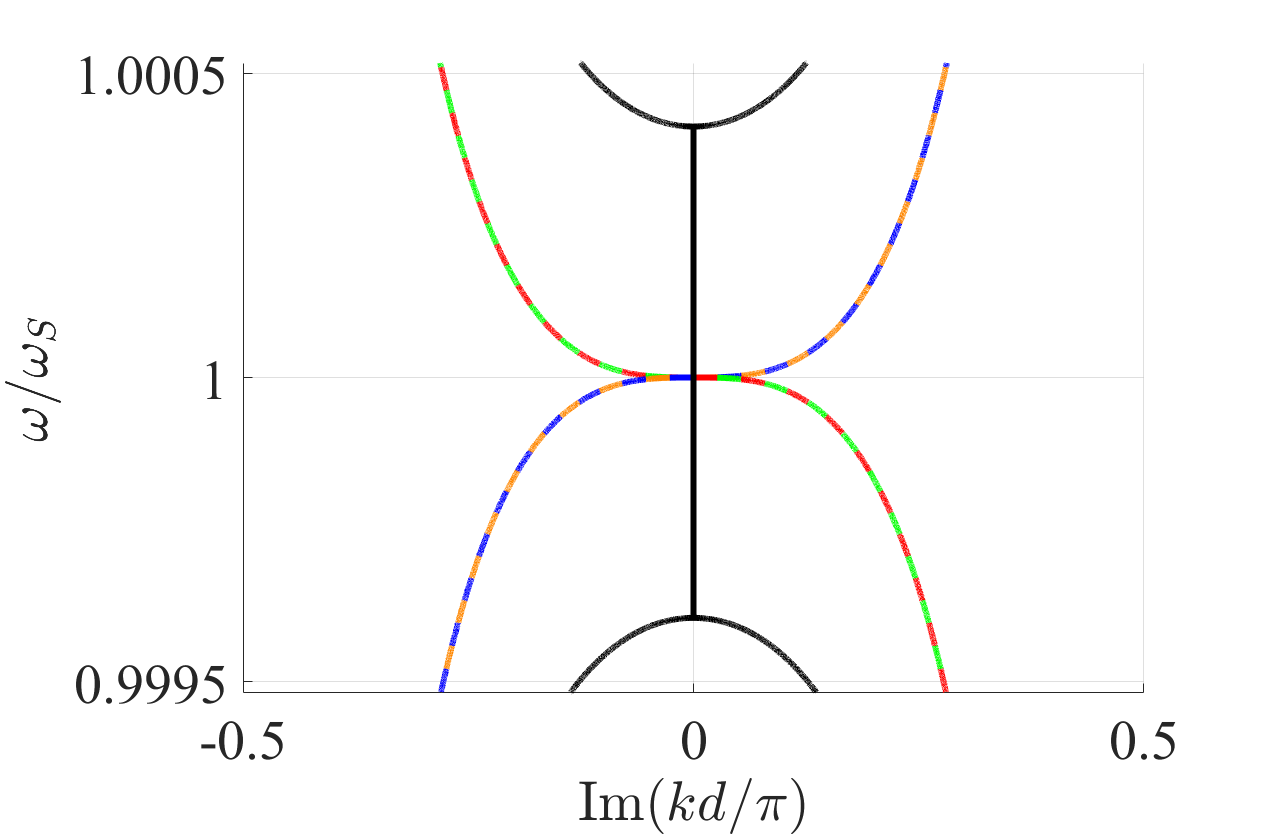}
  \caption{}
 \end{subfigure}
 \medskip
 \begin{subfigure}[t]{.47\textwidth}
  \centering
  \includegraphics[width=\linewidth]{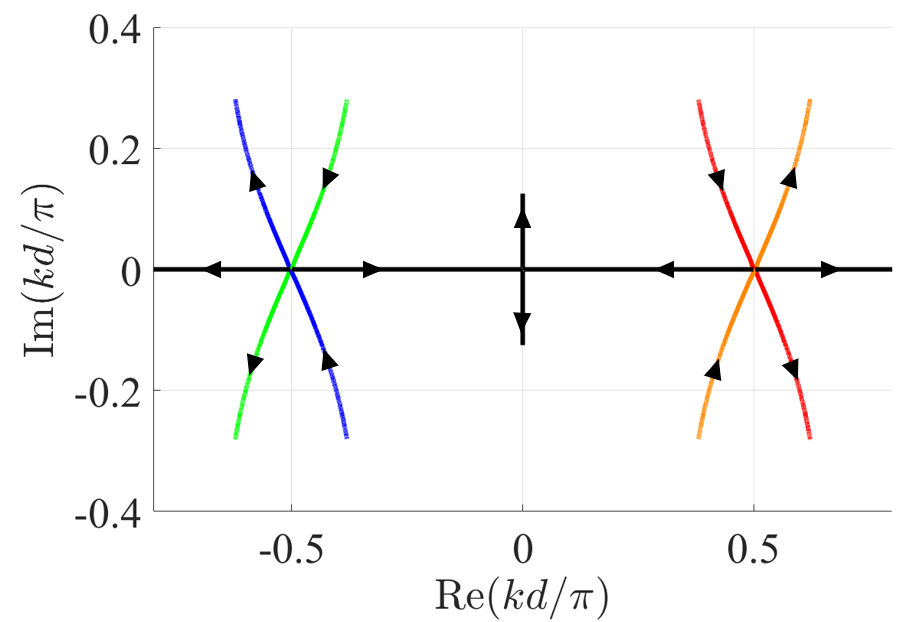}
  \caption{}
 \end{subfigure}
 \hfill
 \begin{subfigure}[t]{.49\textwidth}
  \centering
  \includegraphics[width=\linewidth]{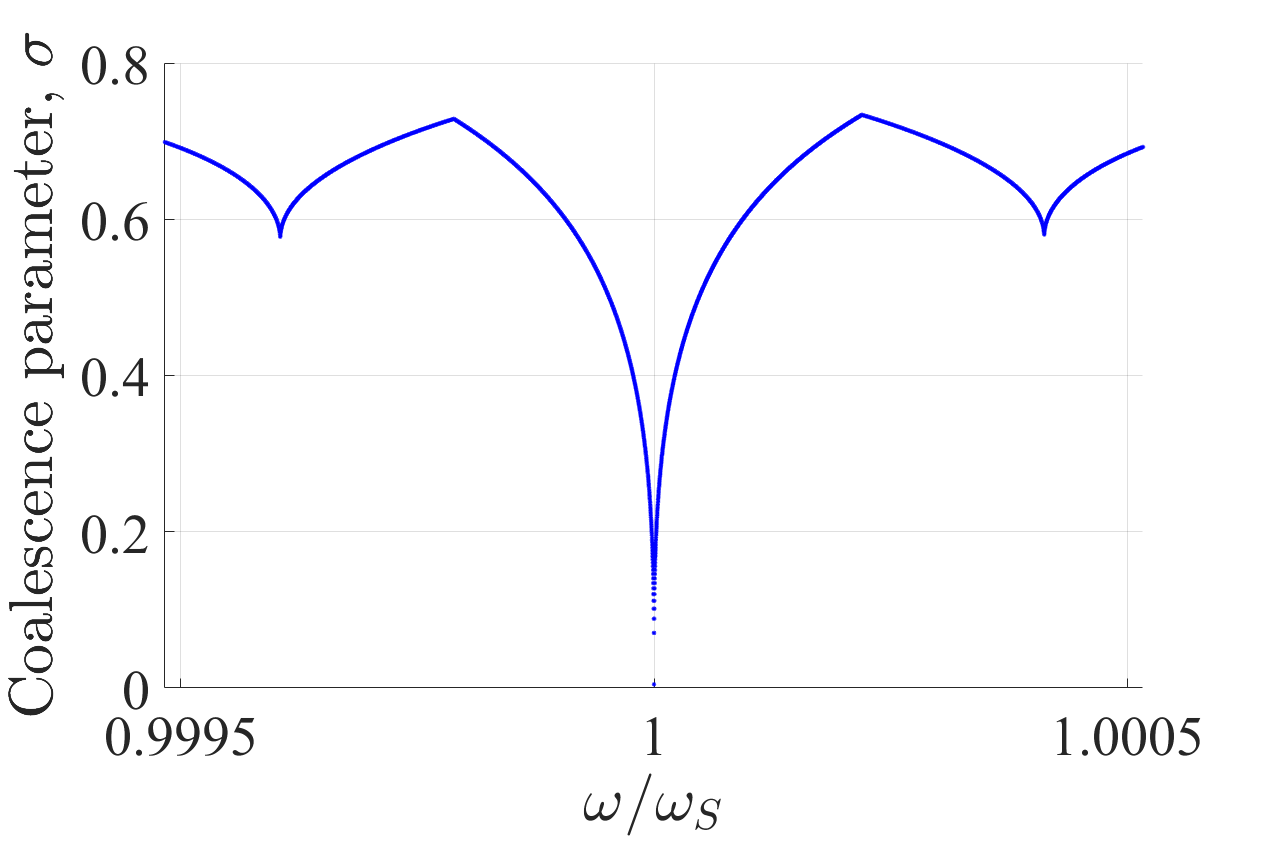}
  \caption{}
 \end{subfigure}
 \caption{Modal dispersion diagram of the lossless and gainless SIP-ASOW. (a) The real part of the eigenvalue versus angular frequency in the fundamental BZ. Solid black: modes with purely real $k$; dashed colors: modes with complex $k$ (overlapping dashed colors imply two overlapping branches). Three curves meet at an inflection point, with reciprocal $k_S$ and $-k_S$ positions. (b) The imaginary part of $k$ versus angular frequency. At the SIP, the three degenerate modes have $\text{Im}(k)=0$. The black curve has $\text{Im}(k)=0$ also near the SIP frequency. (c) Alternative representation of the dispersion diagram in the complex $k$ space. The arrows point in the direction of increasing frequency. (d) Coalescence parameter $\sigma$ versus angular frequency. The coalescence parameter $\sigma$ vanishes at an SIP. The local minima of $\sigma$ indicate the presence of RBEs near the SIP.}
 \label{fig:SIPDispDiagr}
\end{figure}

and its ``evolution'' from unit cell to cell is given by $\boldsymbol{\text{$\psi$}}(n)=\underline{\mathbf{T}}_u \boldsymbol{\text{$\psi$}}(n-1)$, where $\underline{\mathbf{T}}_u$ is the transfer matrix of the unit cell, given in \cite{parareda_frozen_2021}. The time convention $e^{j\omega t}$ is adopted in this paper. The ASOW is a reciprocal structure; it supports six independent modes (unless an EPD occurs) that appear in symmetric positions with respect to the center of the Brillouin zone (BZ) \cite{nada_theory_2017}. The unit cell is engineered so the three modes with the same sign of $\text{Re}(k)$ coalesce at the SIP angular frequency $\omega_S$ \cite{parareda_frozen_2021}. The occurrence of the SIP (which is an EPD of order three) is demonstrated by the vanishing of the coalescence parameter $\sigma$, whose concept was introduced in \cite{abdelshafy_exceptional_2019} for a DBE (where the authors referred to it as a figure of merit or hyperdistance), and applied to an SIP in Refs.~\cite{nada_frozen_2021,parareda_frozen_2021}. In Fig.~\ref{fig:SIPDispDiagr} we show the modal dispersion diagram of an SIP in ASOW, an example which is referred to throughout the paper as the SIP-ASOW. Its SIP frequency is $f_S = 193.54$~$\text{THz}$ and it has structure parameters: $R = 6$ $\upmu \text{m}$, $\alpha_1 = 67.4 ^{\circ} $, $\alpha_2 = 55.6 ^{\circ} $ and $\kappa = 0.5$. The ASOW is made of a rectangular waveguide with a width of $450\;\mathrm{nm}$ and a height of $220\;\mathrm{nm}$, whose modal effective refractive index is $n_w=2.36$ assumed constant in the frequency range of interest \cite{parareda_frozen_2021, scheuer_serpentine_2011}. 

Figures~\ref{fig:SIPDispDiagr}(a) and (b) depict the real and the imaginary parts of the wavenumber against the angular frequency in the fundamental BZ. Solid black lines represent modes with purely real $k$ whereas dashed colors denote modes with complex $k$. The propagating modes (black solid line) in Figures~\ref{fig:SIPDispDiagr}(a) are the same ones shown also in Fig.~\ref{fig:Fig1Waveguide}(b). At the SIP angular frequency $\omega_S$, the wavenumbers of the three modes coalesce at the two inflection points in reciprocal positions with respect to the center of the BZ, and the imaginary part of all modes vanishes. Figure~\ref{fig:SIPDispDiagr}(c) shows the dispersion diagram in the complex $k$ space, where the arrows point in the direction of increasing frequency. The three wavenumbers coalescence is very clear in this figure, and at each SIP the angles between the curves are 60$\degree$. Figure~\ref{fig:SIPDispDiagr}(d) shows the coalescence parameter $\sigma$, defined as in \cite{parareda_frozen_2021}, versus angular frequency. It vanishes at the SIP frequency, clearly showing the coalescence of three eigenvectors and therefore indicating the occurrence of an EPD of order three (the SIP). The two local minima indicate the presence of RBEs at frequencies near the SIP. While the transfer matrix $\underline{\mathbf{T}}_u$ of the unit cell is non-Hermitian and J-unitary at any frequency (see Refs.~\cite{othman_theory_2017, abdelshafy_exceptional_2019, figotin_gigantic_2005}), at the SIP frequency, $\underline{\mathbf{T}}_u$ is similar to a Jordan canonical matrix; it cannot be diagonalized \cite{figotin_frozen_2006, nada_theory_2017}. The EPD in this paper (the SIP) is found in a lossless and gainless ASOW, demonstrating that the presence of gain and loss is not necessary to produce an EPD. 

The three descriptors of an SIP in a loaded finite-length ASOW of $N$ unit cells are the transfer function $T_f$, the group delay $\tau_g$, and the quality factor $Q$. The finite-length ASOW is shown in Fig.~\ref{fig:Fig1Waveguide}(a) where the first and last unit cells are terminated on straight waveguides, without any mirrors and any discontinuity in the waveguide, using Port $1$ on either the left or the right sides of the unit cell shown in Fig.~\ref{fig:Fig2UnitCell}. However, since each unit cell is defined with three ports on each side, Ports $2$ and $3$ on either end of the finite-length ASOW are terminated without coupling to adjacent cells, as in \cite{parareda_frozen_2021}. The transfer function is defined as

\begin{equation}
  T_{f}(\omega)=\frac{E_{out}}{E_{inc}}=\frac{E_{1}^+(N)}{E_{1}^{+}(0)},
  \label{eq:Transfer Function}
\end{equation}

where $E_1^+(0)$ is the incident signal and $E_1^+(N)$ is the transmitted wave onto the attached straight waveguide on the right side of Fig.~\ref{fig:Fig1Waveguide}(a). The group delay is \cite{othman_giant_2016},

\begin{equation}
  \tau_g(\omega) = -\frac{\partial \angle T_f(\omega)}{\partial \omega}
  \label{eq:GroupDelay}
\end{equation}

where $\angle T_f(\omega)$ is the phase of the transfer function. The quality factor for large $N$ is reliably approximated by \cite{nada_theory_2017}

\begin{equation}
  Q = \frac{1}{2}\tau_g(\omega_{S,res})\omega_{S,res},
  \label{eq:QDef}
\end{equation}

where $\omega_{S,res}$ is the resonance frequency closest to the SIP frequency $\omega_S$, referred to as the SIP resonance. For long lossless-gainless periodic waveguides, the asymptotic trend is that $Q\propto N^3$ \cite{figotin_slow_2011, parareda_frozen_2021}. From Eq.~(\ref{eq:QDef}), the group delay at the SIP resonance also scales asymptotically as $\tau_g(\omega_{S,res}) \propto N^3$. The concept of transfer function is used later on to determine the lasing threshold.

\section{Lasing theory in coupled waveguides with SIP}
\label{ch:LasingTheory}

\subsection{Steady-state gain medium response}

Lasing in the lossless ASOW requires doping it with a gain medium, e.g. Erbium, or with other methods, such as using quantum wells. We assume that the gain bandwidth of the active medium includes the SIP frequency of the underlying passive structure. Assuming the gain is uniformly distributed along the waveguide, the mode complex effective refractive index is \cite{yariv_photonics_2007}

\begin{equation}
  n = n_w - jn_g^{\prime\prime},
  \label{eq:ComplexRefrIndex}
\end{equation}

where the subscript $g$ stands for gain and $^{\prime\prime}$ indicates the number has an imaginary value. In the first-order approximation, we neglect the effect of gain on the real part of the mode effective refractive index within the frequency range of interest. The conversion between gain modeled as an imaginary part of the mode effective refractive index $n_g^{\prime\prime}$ and as the gain coefficient $\alpha_g$ (in units of $\text{dB}/\text{cm}$) at an angular frequency $\omega$ is $n_g^{\prime\prime}(\omega) = -\alpha_g (100/8.686)(c/\omega)$, where $c$ is the speed of light in $\text{m/s}$. The imaginary part of the mode effective refractive index is negative for gain, i.e., $n_g^{\prime\prime} <0$.

\subsection{One-pass amplification in ASOW}

Waves traveling in an ASOW at a frequency close to $\omega_S$ experience a high group delay \cite{parareda_frozen_2021}, which increases the effective length of the cavity $L_{eff} = c\tau_g$ \cite{othman_giant_2016, verdeyen_laser_1995}. In turn, this increases the effective power gain coefficient $g_{eff}= \gamma L_{eff}$ in active devices, where $\gamma = -2k_0n_g^{\prime\prime}f$ is the per-unit-length power gain coefficient \cite{yariv_photonics_2007}. The term $f$ is the filling factor of the active material in the host medium. Therefore, operating the active ASOW at a resonance near $\omega_S$ results in an enhanced effective power gain coefficient, which is estimated as 

\begin{equation}
  g_{eff} \approx -2\omega_{res}n_g^{\prime\prime}\tau_g (\omega_{res})f,
  \label{eq:geff}
\end{equation}

and the one-pass effective power gain as

\begin{equation}
  G_{eff} \approx |T_f(\omega_{res})|^2e^{g_{eff}},
  \label{eq:Geff}
\end{equation}

where the $T_f$ and the $\tau_g$ (used to calculate $g_{eff}$) belong to the lossless-gainless finite-length ASOW. This approximation is valid for small values of gain for which the modal properties of the SIP are not significantly perturbed \cite{othman_giant_2016,grgic_fundamental_2012}. Section~\ref{sec:gaindistort} provides a more in-depth discussion on the effects of gain and loss on the distortion of the modes associated with an SIP by examining its effect on the dispersion diagram and on the coalescence parameter of the waveguide modes. The effective power gain coefficient $g_{eff}$ is rewritten as

\begin{equation}
  g_{eff} = -4n_g^{\prime\prime} Q f,
  \label{eq:AlternativeGainQ}
\end{equation}

which shows why high-$Q$ lasers, such as those operating near an SIP \cite{parareda_frozen_2021}, require less gain than lasers with lower $Q$ factors to provide the same output.

\subsection{SIP lasing threshold scaling}
\label{ch:SIPLasginThreshold}

We define the lasing threshold as the minimum gain required to maintain oscillations within a cavity. In a finite-length waveguide operating at a resonance and with a set gain $n_g^{\prime\prime} = n_{th}^{\prime\prime}$, the corresponding value of the effective power gain coefficient $g_{eff}(\omega_{res})$ is calculated using Eq.~(\ref{eq:geff}). Since the $g_{eff}$ parameter is the actual gain parameter that determines the system to start oscillations, when we set this to assume the threshold value, from Eq.~(\ref{eq:AlternativeGainQ}) one infers that the refractive index gain threshold $n_{th}^{\prime\prime} \propto 1/Q $. This relationship indicates a trend, because this discussion (as the one in the previous section, where Eq.~(\ref{eq:AlternativeGainQ}) is derived) neglects the fact that the dispersion diagram, hence $\tau_g$ and $Q$, is affected by the presence of gain \cite{othman_giant_2016,grgic_fundamental_2012}.

\begin{figure}
\centering
  \includegraphics[width = 0.8\textwidth]{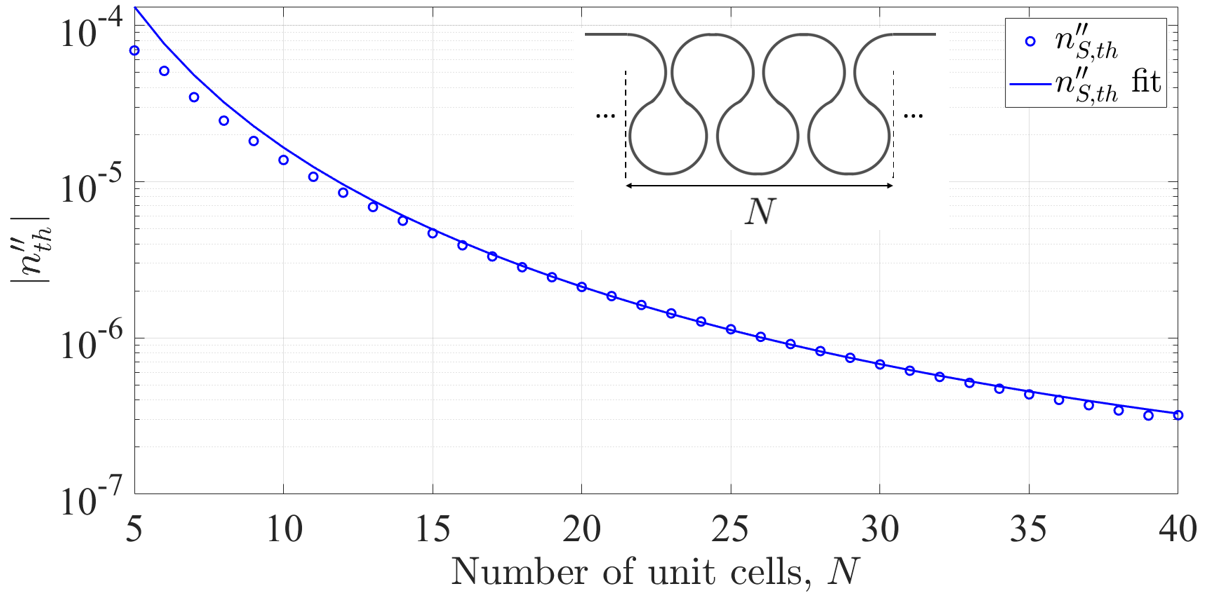}
\caption[\linewidth]{Magnitude of the SIP lasing threshold $|n_{S,th}^{\prime\prime}|$ of a finite-length SIP-ASOW of length $N$. The SIP lasing threshold is fit by $n_{th}^{\prime\prime} = -0.016\times N^{-3} + 7 \times 10^{-8}$, calculated for $N \in [5,40]$. The inset shows a finite-length ASOW of $N$ unit cells terminated without any discontinuity on straight waveguides. The cavity has no mirrors.} 
\label{fig:SIP_nthvsN}
\end{figure}

For asymptotic gainless and lossless periodic waveguides operating near an SIP, $Q \propto N^3$ \cite{figotin_slow_2011, parareda_frozen_2021}. Hence, the SIP lasing threshold, which is calculated at the SIP resonance $\omega_{S,res}$ of an ASOW of finite length, satisfies $n_{S,th}^{\prime\prime} \propto N^{-3}$ asymptotically. Figure~\ref{fig:SIP_nthvsN} depicts this trend for the lossless SIP-ASOW, where the SIP lasing threshold is represented by blue dots. The blue line is the fitting curve $n_{th}^{\prime\prime} = -0.016\times N^{-3} + 7 \times 10^{-8}$, calculated for $N \in [5,40]$. We calculate the SIP lasing threshold with the approach described in Appendix~A. The inversely proportional relationship between the quality factor and the lasing threshold in a conventional cavity \cite{doronin_universal_2021} has been observed having different scaling laws near RBEs ($n_{th}^{\prime\prime}\propto N^{-3}$), DBEs ($n_{th}^{\prime\prime}\propto N^{-5}$) in photonic crystals \cite{othman_giant_2016} and waveguides \cite{veysi_degenerate_2018}, and $6$DBEs ($n_{th}^{\prime\prime}\propto N^{-7}$) in periodic waveguides \cite{nada_exceptional_2020}. In comparison, here we have demonstrated that when working near the SIP, the lasing threshold scaling is $n_{th}^{\prime\prime}\propto N^{-3}$. Note that cavities made of waveguides possessing the SIP have no mirrors at their left and right ends; the cavity effect is due to the "structured" degenerate resonance where the energy distribution vanishes at the cavity edges due to the frozen mode regime, as was shown in \cite{othman_giant_2016} for the DBE and in \cite{ballato_frozen_2005} for the SIP.

\section{Practical considerations}
\label{ch:PracticalConsiderations}

\subsection{Distributed losses}

We consider distributed losses and analyze their effect on the lasing threshold of an active ASOW. Radiation losses due to the waveguide bends in the ASOW are modeled as a positive imaginary part $n_r^{\prime\prime}$ of the mode effective refractive index, leading to

\begin{equation}
  n = n_w - j\left(n_g^{\prime\prime} + n_r^{\prime\prime}\right).
  \label{eq:ImagRefrIndexGainLoss}
\end{equation}

Based on the full-wave simulations in Ref.~\cite{Mealy_CrowSIP2022}, bending losses are estimated as $n_r^{\prime\prime}=2.3\times 10^{-5}$ for a waveguide with radius $R=6$ $\text{$\upmu$m}$ and $n_r^{\prime\prime}= 4.7 \times 10^{-6}$ for $R=10$ $\text{$\upmu$m}$. Therefore, the magnitude of the lasing threshold $|n_{L,th}^{\prime\prime}|$ of the lossy ASOW is 

\begin{equation}
  |n_{L,th}^{\prime\prime}| = |n_{th}^{\prime\prime}| + n_{r}^{\prime\prime},
  \label{eq:LasingThresholdwithLosses}
\end{equation}

which shows radiation losses shift the curve shown in Fig.~\ref{fig:SIP_nthvsN} upwards. Other distributed losses can also be modeled as a positive imaginary part of the effective refractive index, causing a further shift of the lasing threshold. In the following, we will show that the considered losses and gain, even though they distort the dispersion diagram, are not large enough to fundamentally modify the properties of the frozen mode associated to the SIP, i.e., the coalescence parameter still tends to vanish (Fig. \ref{fig:DistortedDispDiagr}), which explains why the SIP lasing threshold still scales asymptotically as a negative cube of the waveguide length.

\subsection{Gain and loss-induced mode distortion}
\label{sec:gaindistort}

We discuss the modal dispersion diagram distortion induced by the presence of either gain or loss in the otherwise lossless-gainless SIP-ASOW. Eigenmodes at an EPD display exceptional sensitivity when a system parameter is perturbed by a small quantity \cite{figotin_slow_2006}. The insertion of gain or loss in the lossless-gainless ASOW perturbs the mode effective refractive index (i.e., of the quasi-TE mode in the infinitely long homogeneous straight rectangular waveguide), by a small quantity $\Delta n$. Therefore, the degenerate wavenumber $k_S$ at the SIP frequency splits into three wavenumbers $k_q$, with $q = 1,2,3$, estimated by a first-order approximation of the Puiseux fractional power series \cite{figotin_slow_2006, nada_theory_2017, kato_perturbation_1966} as 

\begin{equation}
	k_q(\Delta n) \approx k_S + a_1e^{j\frac{2\pi }{3}q} (\Delta n)^{1/3} ,
	\label{eq:PuiseuxSeries}
\end{equation}

where $a_1$ is a constant calculated following the expressions in Ref.~\cite{welters_explicit_2011}. The wavenumber perturbation at and near the SIP frequency, due to gain or loss, is stronger than the wavenumber perturbation at frequencies away from the SIP. This important phenomenon is understood by comparing the result in Eq.~(\ref{eq:PuiseuxSeries}) with the conventional Taylor expansion for the wavenumber perturbation when the frequency is sufficiently far from the SIP, leading to each of the three wavenumbers (in each direction) $k_q$ (with $q = 1, 2, 3$) at any $\omega$ away from the SIP to be approximated as $k_q (\Delta n) \approx k_{q,0} + b_q (\Delta n)$, where $b_q=\partial k_q / \partial n$ is the standard coefficient of a Taylor expansion when a perturbation $\Delta n$ of the modal refractive index $n$ (of the homogeneous straight waveguide quasi-TE mode) is applied. 
Therefore, when we look at the wavenumber perturbation from the SIP (i.e., when the system operates at the SIP frequency), one has $\Delta k_{SIP}=k_q-k_S \propto (\Delta n)^{1/3}$. This generates a larger wavenumber perturbation than $\Delta k_q=k_q-k_{q,0} \propto (\Delta n)$, i.e., when not working at the SIP frequency. As an example, when $|\Delta n| = 10^{-3}$, i.e., we modify the third decimal digit of the effective refractive index, one has $|\Delta n|^{1/3} = 10^{-1}$ which is much larger than $10^{-3}$. Therefore, when in Eq.~(\ref{eq:ImagRefrIndexGainLoss}) we consider a small $\Delta n = n_g-n_w=- j\left(n_g^{\prime\prime} + n_r^{\prime\prime}\right)$, it is clear that the dispersion diagram at the SIP is more perturbed than at any other frequency. This is confirmed by the following simulations.

\begin{figure}[t]
 \begin{subfigure}[t]{.50\textwidth}
  \centering
  \includegraphics[width=\linewidth]{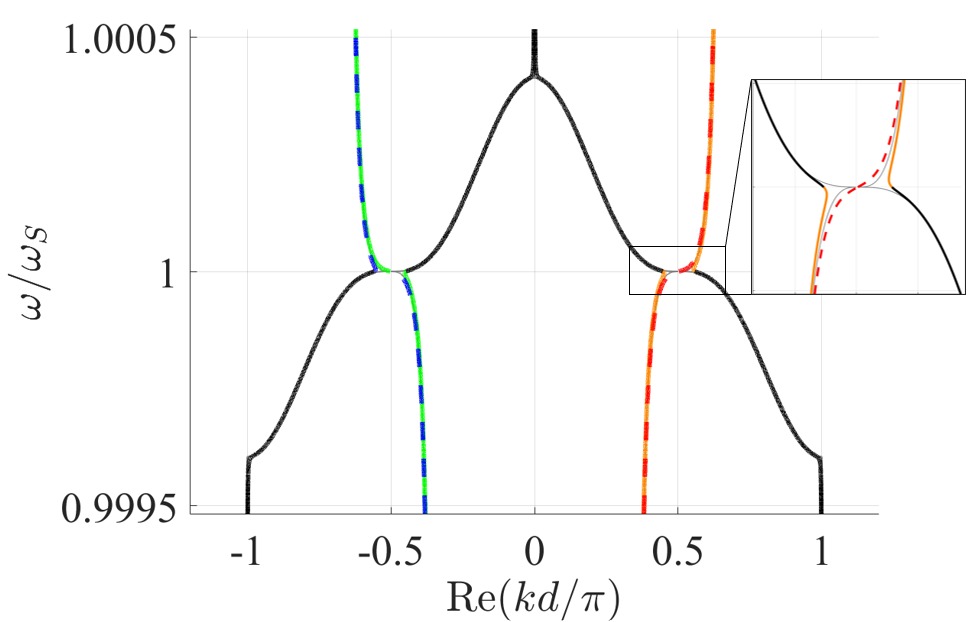}
  \caption{}
 \end{subfigure}
 \hfill
 \begin{subfigure}[t]{.47\textwidth}
  \centering
  \includegraphics[width=\linewidth]{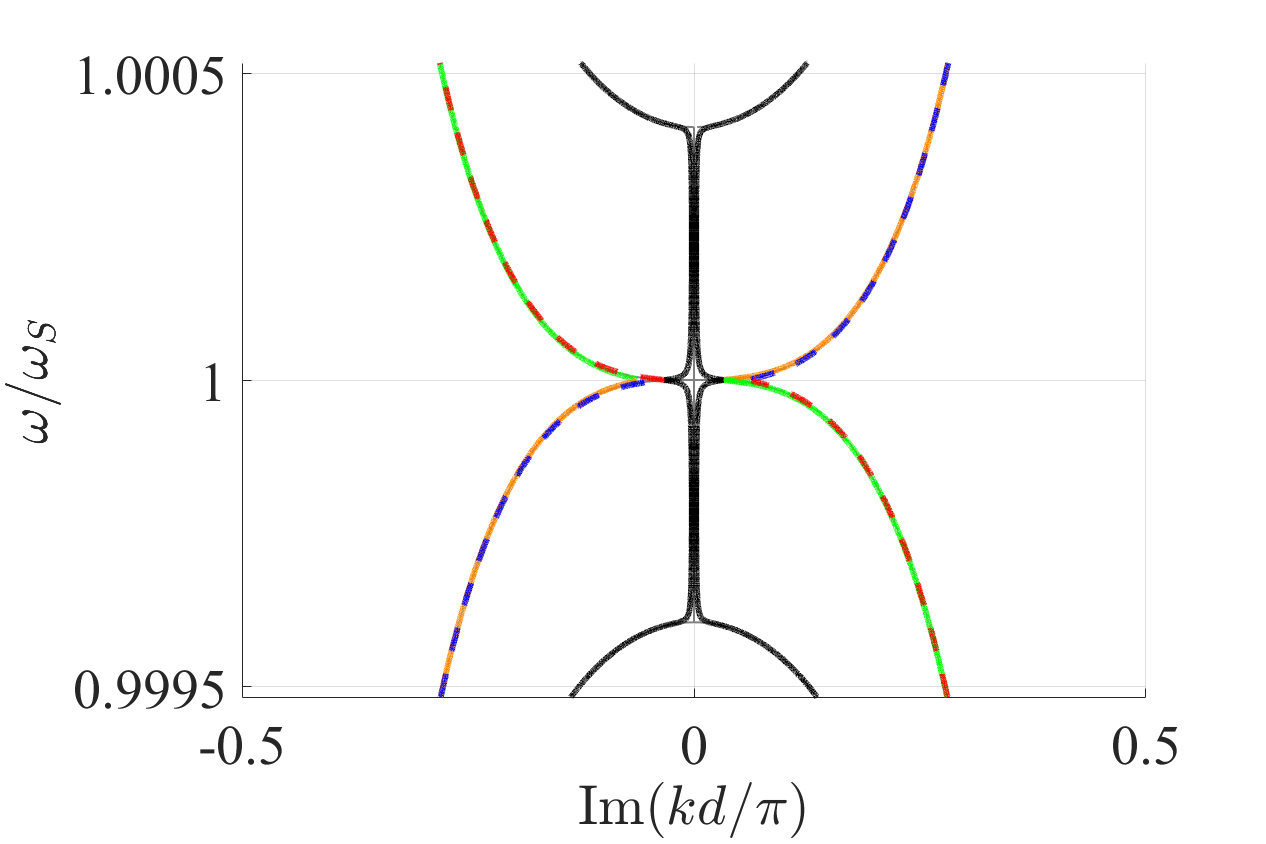}
  \caption{}
 \end{subfigure}
 \medskip
 \begin{subfigure}[t]{.48\textwidth}
  \centering
  \includegraphics[width=\linewidth]{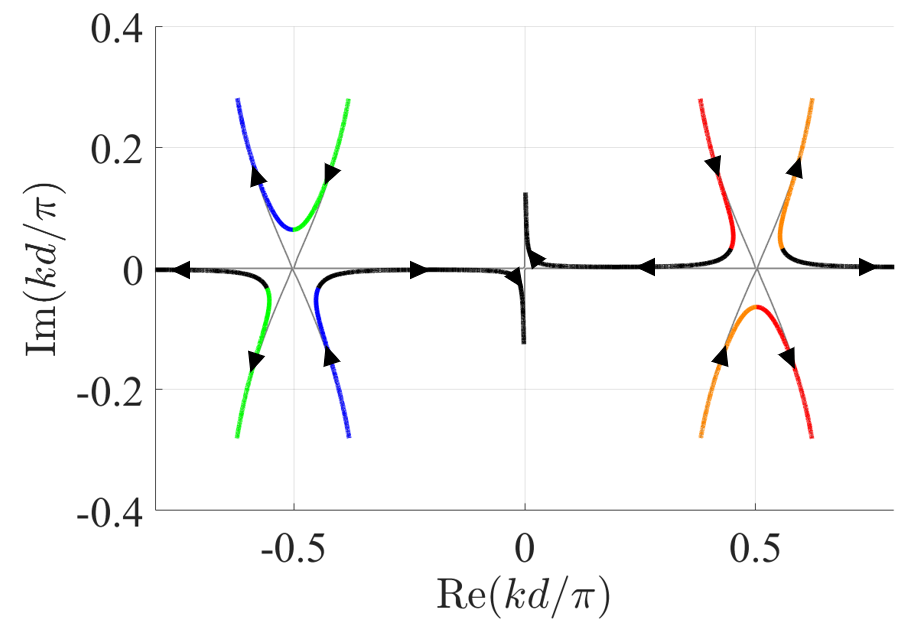}
  \caption{}
 \end{subfigure}
 \hfill
 \begin{subfigure}[t]{.49\textwidth}
  \centering
  \includegraphics[width=\linewidth]{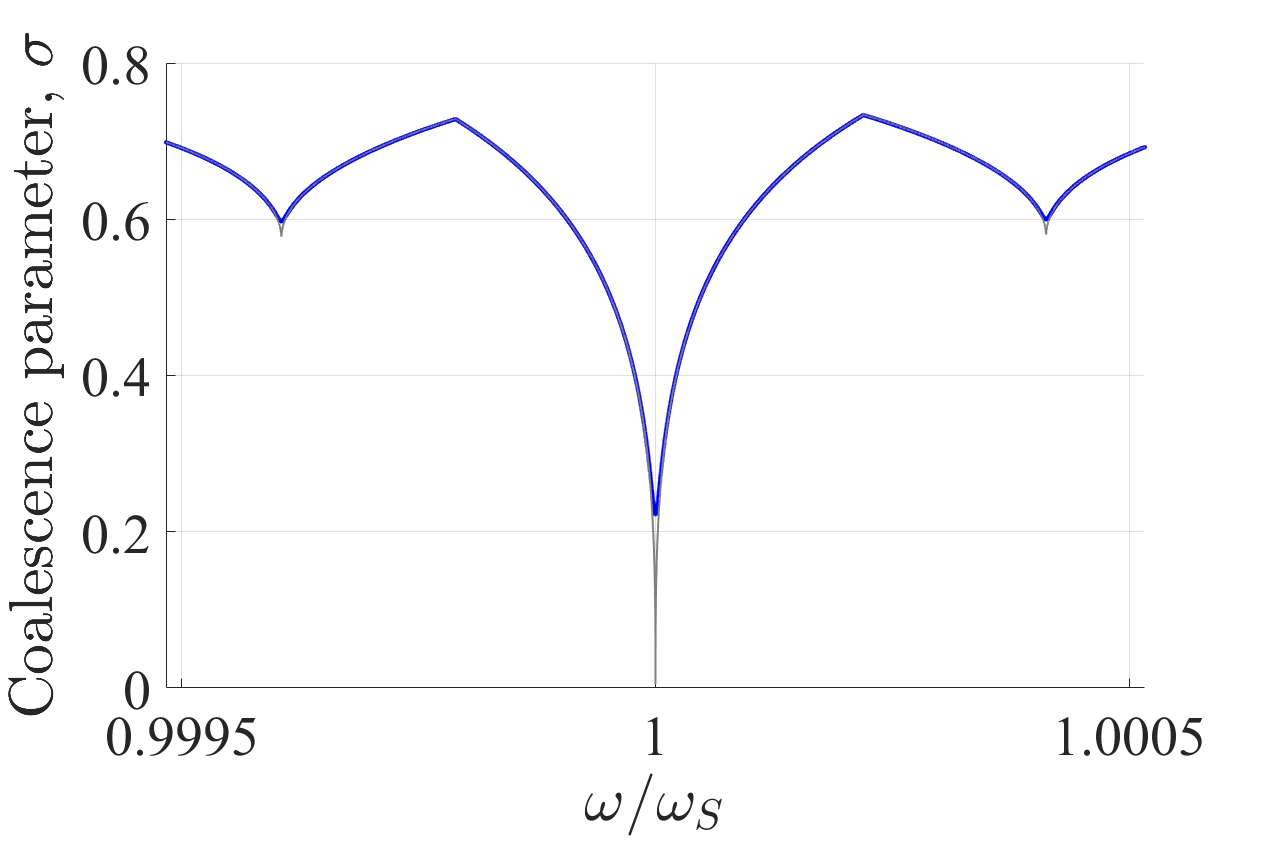}
  \caption{}
 \end{subfigure}
 \caption{Modal dispersion diagram of the SIP-ASOW with a gain of $n_g^{\prime\prime} = -8\times 10^{-6}$, equivalent to $\alpha_g = 2.82$ $ \text{dB}/\text{cm}$. The black curves represent propagating modes, with purely real wavenumbers; the dashed, colored curves represent evanescent modes. The dispersion diagram for the SIP-ASOW without gain is shown in light grey. (a) Real part of the wavenumber versus angular frequency in the fundamental BZ. The inset shows the eigenvalue distortion is accentuated at the SIP. (b) The imaginary part of $k$ versus angular frequency. Due to gain, $\text{Im}(k) \neq 0$ at all the frequencies, especially around $\omega_S$. (c) Alternative representation of the dispersion diagram in the complex $k$ space. The lack of mode coalescence due to gain is very clear in this figure, with the three modes departing from $k_S$ at $120^{\circ}$ from each other as expected from Eq.~(\ref{eq:PuiseuxSeries}). (d) The coalescence parameter $\sigma$ versus angular frequency. The non vanishing $\sigma$ indicates the SIP does not fully form, but the dip still demonstrates a degree of exceptional three-mode degeneracy to be exploited for the SIP laser.}
 \label{fig:DistortedDispDiagr}
\end{figure}

Figure~\ref{fig:DistortedDispDiagr} shows the eigenmode perturbation in a SIP-ASOW without loss but with gain  $n_g^{\prime\prime} = - 8\times 10^{-6}$, equivalent to $\alpha_g = 2.82$ $ \text{dB}/\text{cm}$, superimposed on the dispersion diagram of the same waveguide without gain (thin gray line). Figures~\ref{fig:DistortedDispDiagr}(a) and (b) depict the real and imaginary parts of the wavenumbers of the modes. The inset shows the details of the perturbation of the wavenumber due to gain at the SIP. Any wavenumber perturbation is much less visible at any other frequency. The distortion of the SIP due to the presence of gain is observed more clearly in Fig.~\ref{fig:DistortedDispDiagr}(c), which shows the dispersion diagram in the complex $k$ space. The three wavenumbers do not fully coalesce anymore, and they slightly depart from $k_S$ at angles of $120^{\circ}$ from each other, as shown in Eq.~(\ref{eq:PuiseuxSeries}). Larger gain would cause an even larger departure from $k_S$. These observations are in agreement with what was discussed in \cite{grgic_fundamental_2012} for an RBE and in \cite{yazdi_new_2018} for an SIP: adding gain to the lossless-gainless waveguide deteriorates the degeneracy and its exceptional properties. The extent of the deterioration of the properties, however, remains unclear. As discussed in \cite{grgic_fundamental_2012}, the presence of a small amount of either gain or loss causes a perturbation of the wavenumber (and hence of the group velocity) of the same magnitude.

Despite having explained why the dispersion diagram experiences the largest perturbation in the neighborhood of the SIP, we also show that the EPD-related properties associated to the SIP are however retained. Indeed, with the amount if gain here considered, we still observe the coalescence of the three eigenvectors (i.e., polarization states), that fully describe the degree of the exceptional degeneracy. Indeed, in Fig.~\ref{fig:DistortedDispDiagr}(d) we show that even though the coalescence parameter $\sigma$ does not fully vanish at the SIP, it is still significantly smaller than unity at $\omega_S$. Therefore, the SIP feature of three eigenvectors approaching each other is still present despite the SIP distortion due to gain. Experimental verifications of the existence of an SIP in non-reciprocal and reciprocal waveguides with small losses were reported in \cite{apaydin_experimental_2012, nada_frozen_2021}, where the distorted dispersion diagrams were reconstructed, which resemble the one shown in Fig.~\ref{fig:DistortedDispDiagr}. Despite the distortion, the modes still exhibit properties typically associated with the frozen mode. Therefore, the SIP condition is relatively resistant to the presence of small values of loss or gain. For increasingly large values of gain, however, the group delay at frequencies near the SIP eventually diminishes (because the SIP gracefully loses its properties), and this effect curbs the effective power gain coefficient $g_{eff}$ in Eq.~(\ref{eq:geff}). 

In summary, despite the stronger perturbation of the wavenumbers in proximity of the SIP frequency compared to other frequencies, the coalescence parameter in Fig.~\ref{fig:DistortedDispDiagr}(d) still shows that the three eigenvectors are close to each other, demonstrating a good degree of exceptional degeneracy. On the other hand, larger and larger gain values would strongly perturb the SIP degeneracy till there is no degeneracy anymore.

\section{Lasing at SIP versus RBE}
\label{ch:MultiEPD}

It is typical that optical structures capable of supporting an SIP also have RBEs close to $\omega_S$ \cite{figotin_slow_2011, ramezani_unidirectional_2014, nada_theory_2017, parareda_frozen_2021, paul_frozen_2021, zhi_coupled_2019, tuxbury_scaling_2021}. Tolerances in fabrication might prevent the formation of the SIP, and the laser would then operate near an RBE. The lossless-gainless ASOW is relatively robust to fabrication tolerances because it is formed by a single waveguide with a continuous curvature slope, as seen in Fig.~\ref{fig:Fig2UnitCell}, and therefore some changes would only cause a shift of the SIP frequency. For example, a perturbation in $n_w$ is expected to occur uniformly, so the SIP would still be formed, although at a different frequency. However, some other perturbations may degrade the quality of the SIP (the coalescence parameter would not approach zero), and therefore the properties of the frozen mode might be compromised.

In some scenarios, despite the occurrence of an SIP, a laser might emit at the RBE resonance frequency $\omega_{R,res}$, the closest to the RBE frequency, if it has a lower lasing threshold than the one associated to the resonance at $\omega_{S,res}$.

To prevent RBE lasing, one potential solution is to increase the frequency difference between the SIP and RBE. Consequently, we consider the free spectral range (FSR), defined by the phase accumulated by propagation along the optical length of an ASOW unit cell without coupling (ie: without the frozen mode). From \cite{scheuer_serpentine_2011} adapted to the asymmetric case studied in \cite{parareda_frozen_2021} and here, we have

\begin{equation}
  \Delta f_{FSR} = \frac{c / n_{w}}{2R(\pi + \alpha_1 + \alpha_2)}.
  \label{eq:FSR}
\end{equation}

\begin{figure}[H]
\centering
  \includegraphics[width = 0.8\textwidth]{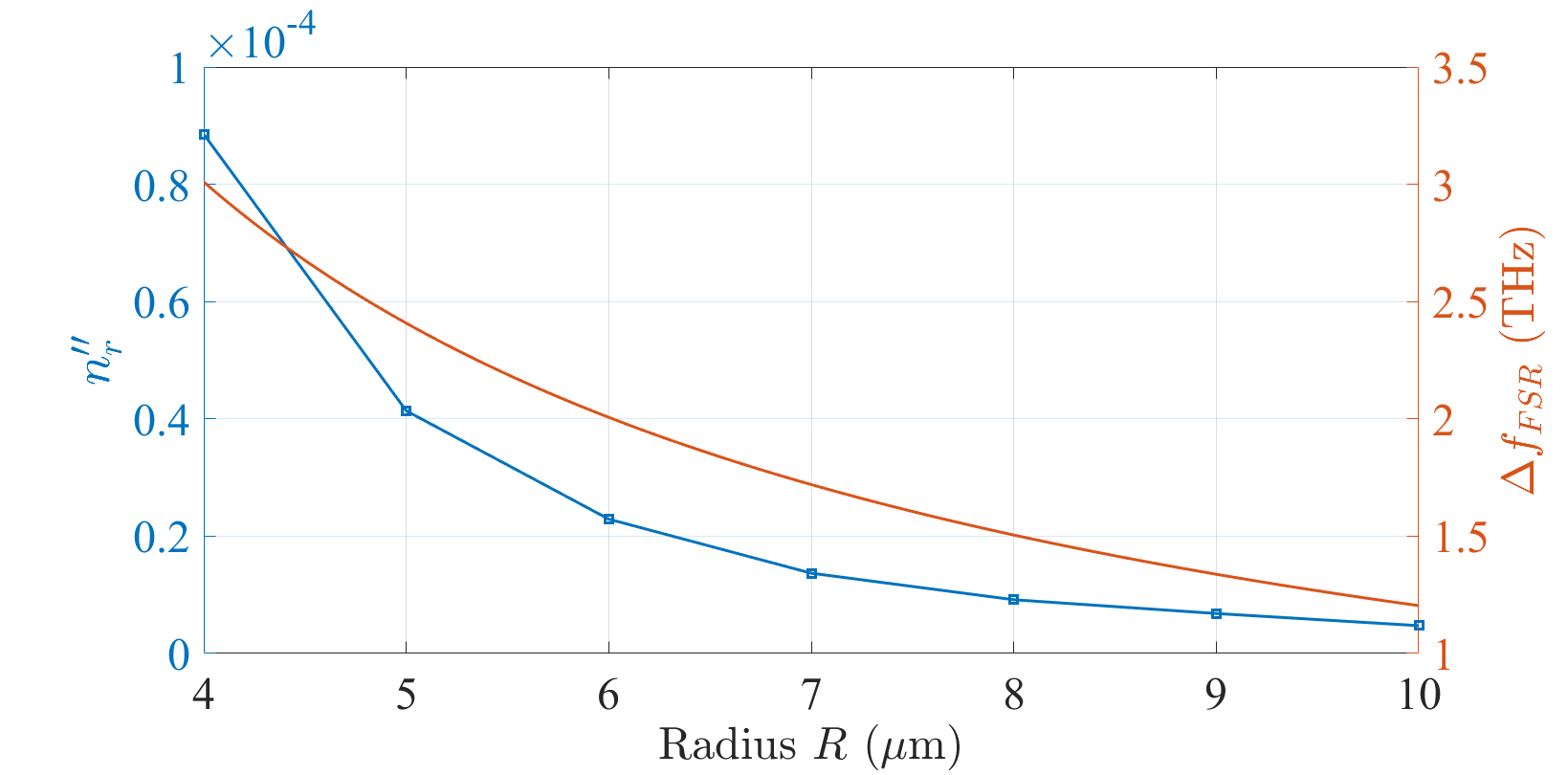}
\caption[\linewidth]{The loss values in terms of the imaginary effective refractive index are extracted from \cite{Mealy_CrowSIP2022}, and the FSR values are calculated using Eq.~(\ref{eq:FSR}) with the structure parameters from the SIP-ASOW. Low radiation losses imply small FSR, which in turns implies a large spectral density of features like SIP and RBE frequencies.}
\label{fig:RadiusLossesFSR}
\end{figure}

Therefore, the distance between the SIP and the RBE is increased by decreasing $R$, which effectively stretches the dispersion diagram. A smaller loop radius, however, is associated with higher radiation losses, as shown in Fig.~\ref{fig:RadiusLossesFSR}. As a compromise, we choose $R=6$ \text{$\upmu$}m in the SIP-ASOW example. In the following discussion, we compare the performance of the SIP-ASOW of different lengths operating either near the SIP or near an RBE (shown in Fig.~\ref{fig:SIPDispDiagr}). We choose to consider the RBE at a lower frequency ($f_R$) than the SIP ($f_S$) because it is the one closest to the SIP (versus the RBE at a higher frequency than the SIP). The frequency difference between these two EPDs is $\Delta f = f_S - f_R = 76.4\; \mathrm{GHz}$.

\begin{figure}[t]
  \centering
  \begin{subfigure}[b]{0.49\textwidth}
    \includegraphics[width =\textwidth]{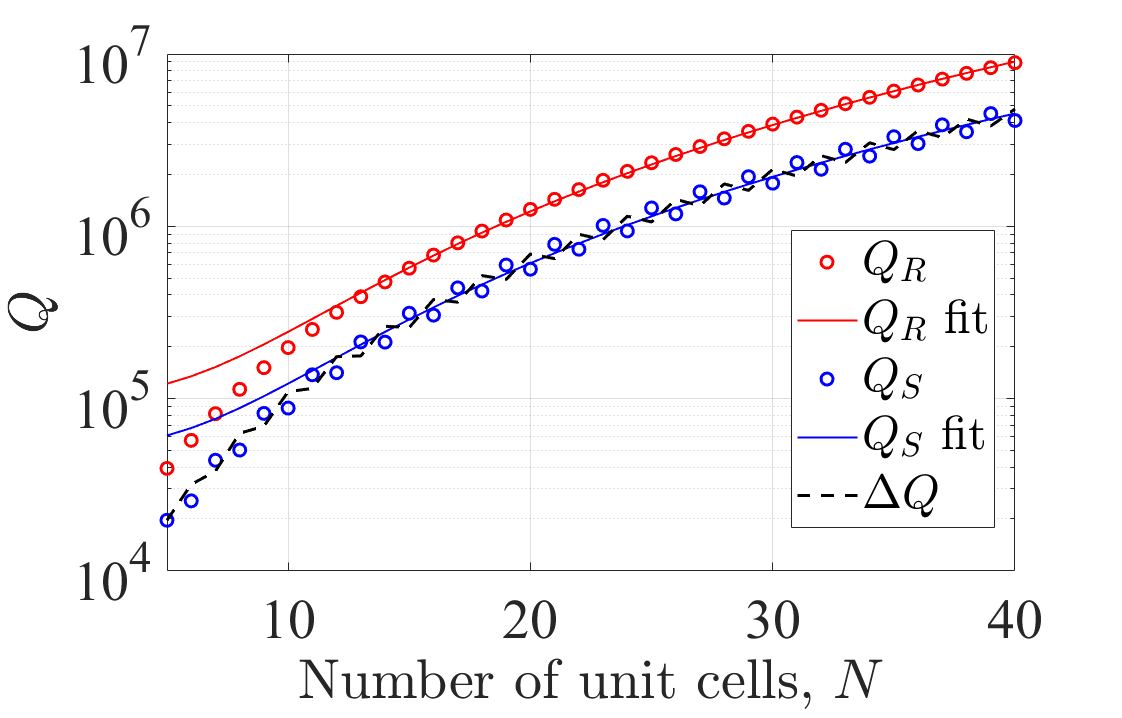}
    \caption{}
    \label{fig:QvsN}
  \end{subfigure}
  \begin{subfigure}[b]{0.49\textwidth}
    \includegraphics[width = \textwidth]{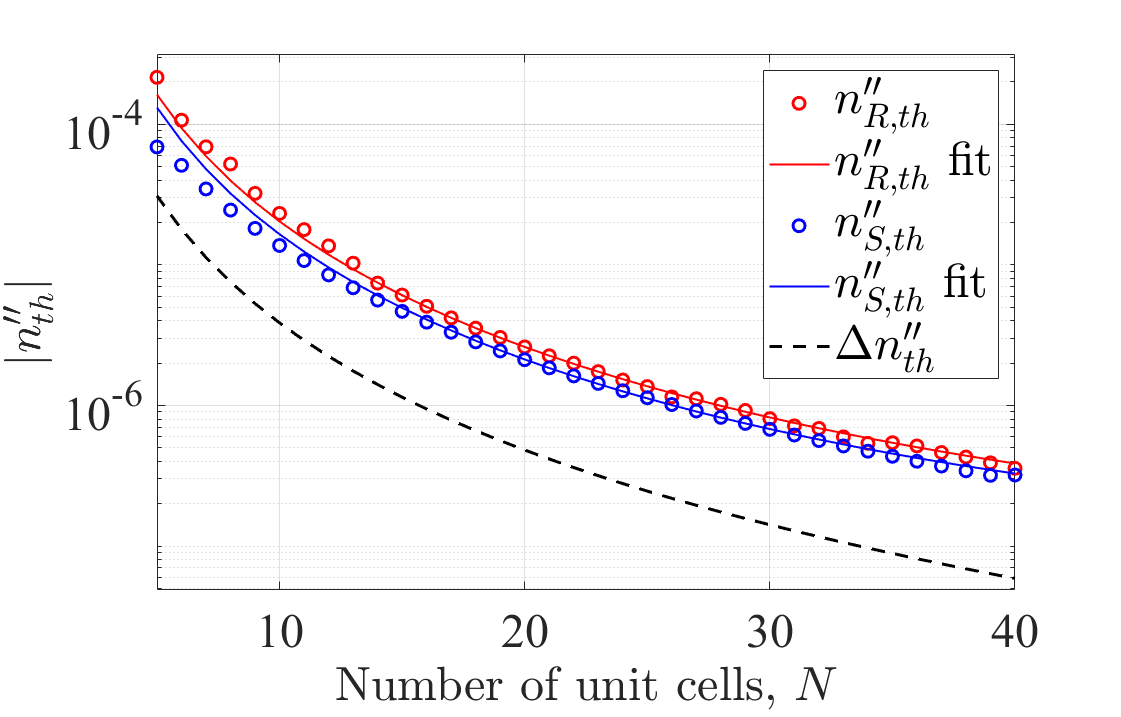}
    \caption{}
    \label{fig:ThresholdvsN}
  \end{subfigure}
  \hfill
  \begin{subfigure}[b]{0.49\textwidth}
    \includegraphics[width =\textwidth]{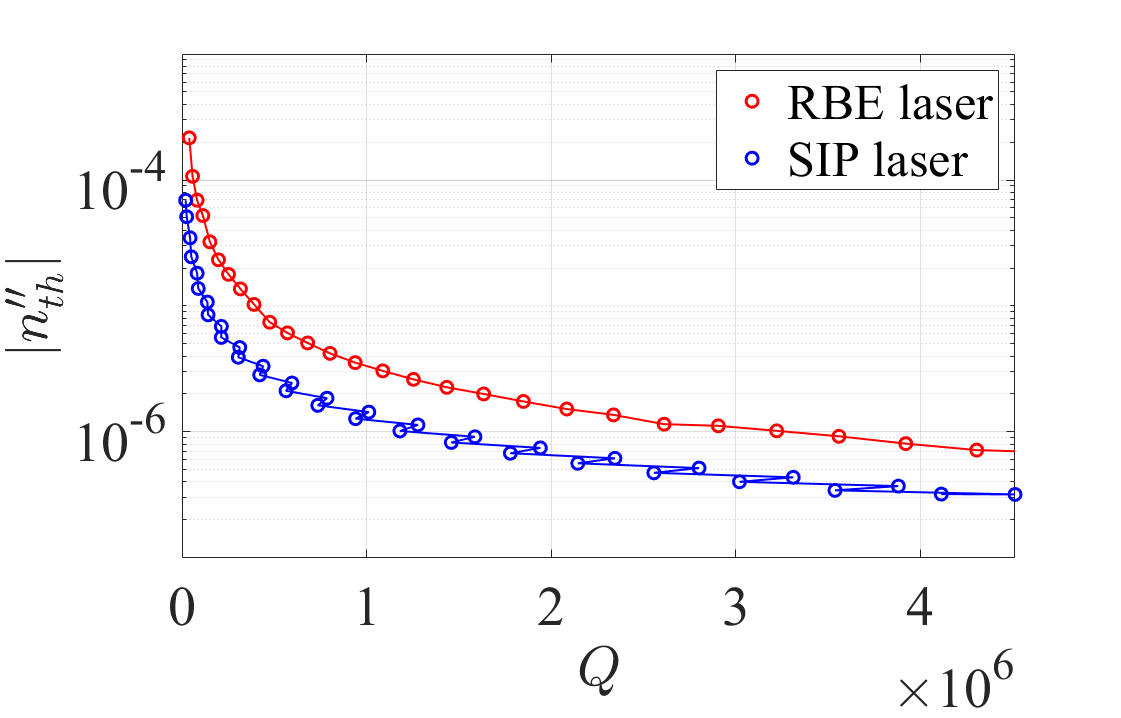}
    \caption{}
  \end{subfigure}
  \caption[\linewidth]{Comparison between the SIP and RBE ASOW lasers of different lengths, operating either near the SIP frequency $\omega_S$ (in blue) or the RBE frequency $\omega_R$ (in red), respectively. (a) The RBE quality factor $Q_R$ (represented by red dots), fitted by the red curve from the first equation in Eq.~(\ref{eq:SIPRBEQfitting}); the SIP quality factor $Q_S$ (represented by blue dots) is fitted by the blue curve following the second equation in Eq.~(\ref{eq:SIPRBEQfitting}); and their difference $\Delta Q \approx Q_R - Q_S$ is represented as a dashed, black line. The three curves scale as $N^{-3}$. (b) Magnitude of the RBE lasing threshold (represented by red dots), fitted by the magnitude of the red curve from the first equation in Eq.~(\ref{eq:SIPRBELasingThr}); and the magnitude of the SIP lasing threshold (represented by blue dots), fitted by the magnitude of the blue curve following the second equation in Eq.~(\ref{eq:SIPRBELasingThr}); and the magnitude of their difference $\Delta n_{th}^{\prime\prime} = n_{ R,th}^{\prime\prime} - n_{ S,th}^{\prime\prime}$ is represented as a dashed, black line. The three thresholds decrease as $N^{-3}$. The lasing threshold associated to the SIP resonance is smaller than the RBE one for an ASOW of the same length. (c) Magnitude of the lasing threshold versus quality factor for an RBE -based laser operating near $\omega_R$ (in black) and for an SIP -based laser operating near $\omega_S$ (in blue). Each dot (red and blue) represents a finite-length ASOW with a given $Q$ provided by some $N\in [5,40]$. An active ASOW operating near an SIP has a lower lasing threshold and lower quality factor than the same ASOW operating near an RBE.}
  \label{fig:SIPRBElasingvsN}
\end{figure}

Assuming both the SIP and the RBE frequencies fall within the active frequency region of the gain medium, as would be the case for an Erbium-doped waveguide \cite{xu_experimental_2021}, the system may start oscillations at the SIP resonance frequency $f_{S,res}$ or at the RBE resonance frequency $f_{R,res}$. We denote by $n_{S,th}^{\prime\prime}$ and $n_{R,th}^{\prime\prime}$ the two gain values that would start oscillations at those two frequencies, respectively. Lasing near the RBE would occur if the RBE lasing threshold, which is calculated at $\omega_{R,res}$ following the method described in Appendix~A, satisfies $|n_{R,th}^{\prime\prime}| < |n_{S,th}^{\prime\prime}|$. Following the discussion from Section~\ref{ch:SIPLasginThreshold}, one would expect lasing near the RBE if the quality factor near the RBE is larger than near the SIP. Figure~\ref{fig:SIPRBElasingvsN}(a) shows the RBE quality factor $Q_R$, calculated using Eq.~(\ref{eq:QDef}) at $\omega_{R,res}$, for different $N$ (in red dots), and the SIP quality factor $Q_S$, calculated at $\omega_{S,res}$ (in blue dots). Both quality factors are calculated at a loaded ASOW, continued on the rectangular waveguides on the left and right sides without mirrors, as in the inset in Fig.~\ref{fig:SIP_nthvsN}. We also plot the fitting curves (solid lines) governed by 

\begin{equation}
\begin{split}
  Q_S &\approx a N^{3} + b, \\
  Q_R &\approx c N^{3} + d, \\
  \Delta Q &\approx Q_R - Q_S = (c-a) N^{3} + (d-b),
\end{split}
  \label{eq:SIPRBEQfitting}
\end{equation}

with coefficients $a = 69.8$, $b = 5.2\times 10^4$, $c=139.6$, and $d=10^5$. As expected, $Q_S$ and $Q_R$ scale asymptotically as $N^3$, when the cavity resonant frequency is near the SIP \cite{figotin_slow_2011, parareda_frozen_2021} and near the RBE \cite{nada_theory_2017}, respectively. Their difference $\Delta Q = Q_R-Q_S$ is shown as a dashed black line, also follows the trend $\Delta Q \propto N^3$ for large $N$, and for the specific SIP-ASOW under consideration, $Q_S << Q_R$. This shows that the SIP-associated frozen mode regime in the current design is not as mismatched to its termination impedances as the cavity associated with the RBE resonance. The small zig-zag distribution of $Q_S$ with waveguide length is discussed in \cite{parareda_frozen_2021}.

Because of the $Q_R\propto N^3$ asymptotic scaling, we expect the RBE lasing threshold to also scale asymptotically as $n_{R,th}^{\prime\prime} \propto N^{-3}$. Indeed, that is what we observe. Fig.~\ref{fig:SIPRBElasingvsN}(b) shows $|n_{R,th}^{\prime\prime}|$ in red dots, fitted by the red solid-line curve, and $|n_{S,th}^{\prime\prime}|$ in blue dots, which is fitted by the blue curve given by 
 
\begin{equation}
\begin{split}
  n_{S,th}^{\prime\prime} &\approx e N^{-3} + f, \\
  n_{R,th}^{\prime\prime} &\approx g N^{-3} + h, \\
  \Delta n_{th}^{\prime\prime} &= n_{R,th}^{\prime\prime} - n_{S,th}^{\prime\prime} \approx (g-e)N^{-3} + (h-f),
\end{split}
  \label{eq:SIPRBELasingThr}
\end{equation}

with coefficients $e = -0.016$, $f = 7\times 10^{-8}$, $g = -0.02$, and $h = 6.9\times 10^{-8}$. The magnitude of the difference between the lasing thresholds for lasers working at the RBE resonance or at the SIP resonance, $\Delta n_{th}^{\prime\prime} = n_{R,th}^{\prime\prime} - n_{S,th}^{\prime\prime}$, is depicted as a dashed black line. The difference in lasing thresholds between the SIP and RBE also scales asymptotically as $N^{-3}$, resulting in comparable thresholds for large waveguides. Therefore, to balance the preservation of the frozen mode properties with the prevention of accidental lasing near an RBE, the waveguide length should be chosen as a trade-off between achieving a small $|n_{S,th}^{\prime\prime}|$ and a relatively large $\Delta n_{th}^{\prime\prime}$.

\begin{figure}[t]
  \centering
  \begin{subfigure}[b]{0.49\textwidth}
    \includegraphics[width = \textwidth]{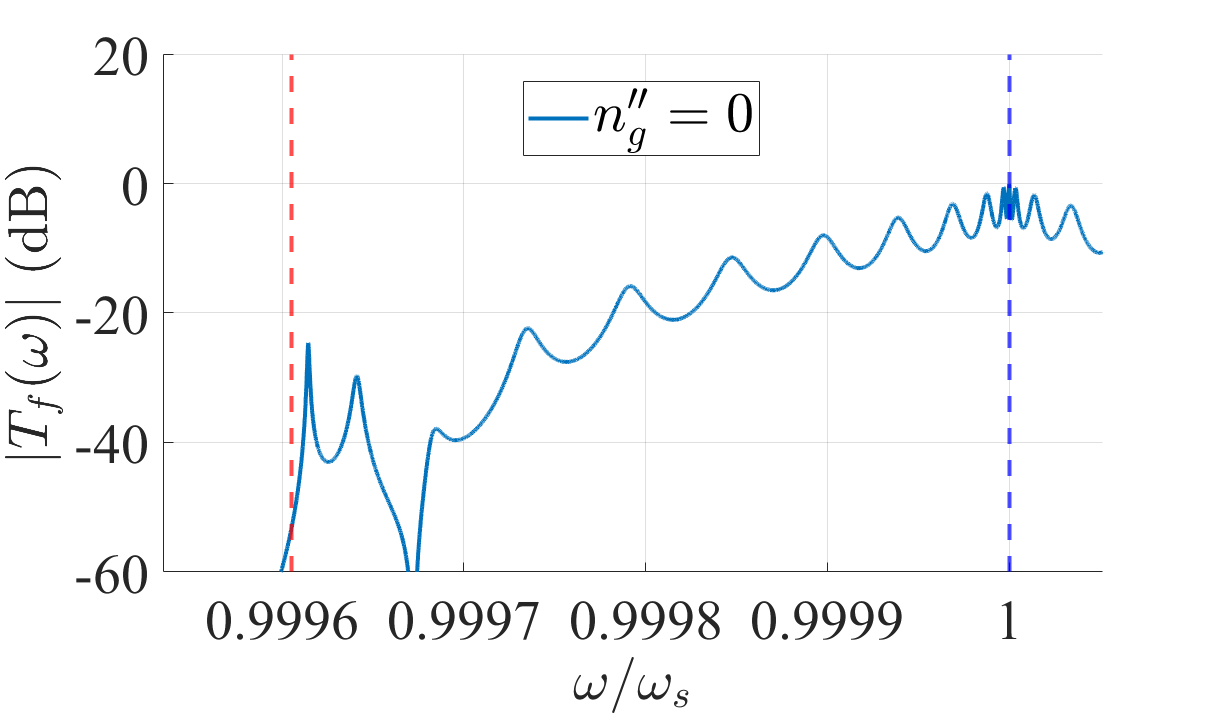}
    \caption{}
  \end{subfigure}
  \begin{subfigure}[b]{0.49\textwidth}
    \includegraphics[width =\textwidth]{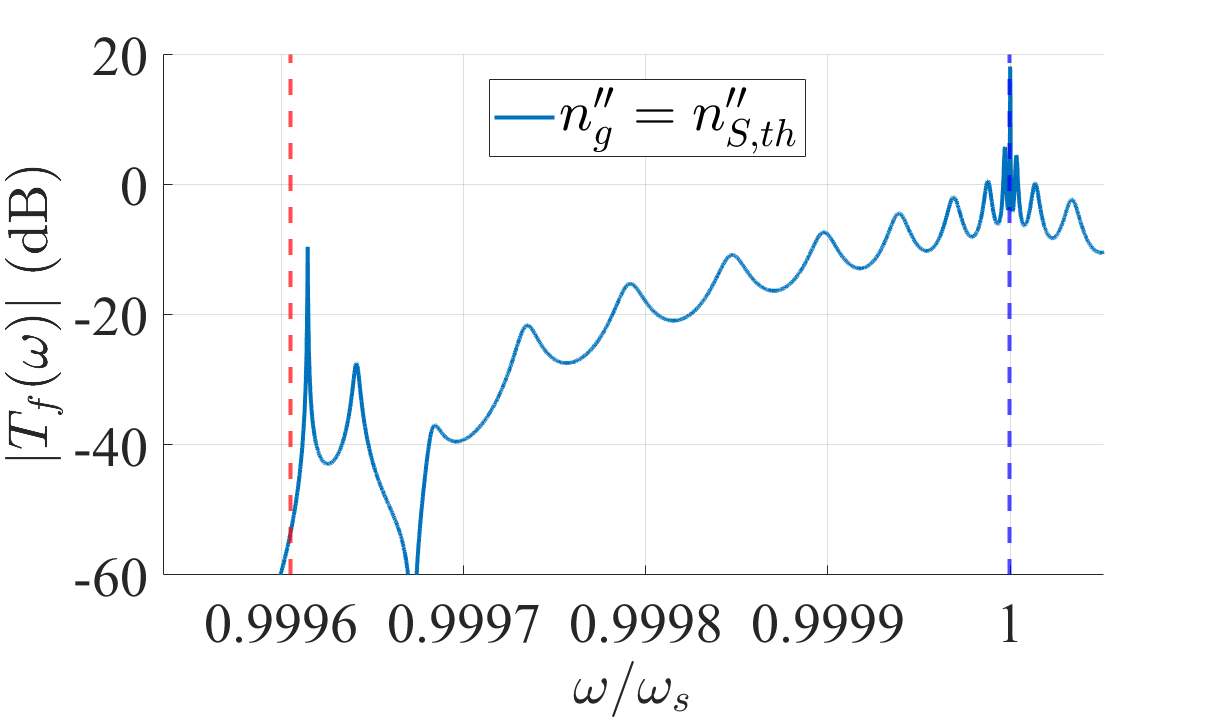}
    \caption{}
  \end{subfigure}
  \hfill
  \begin{subfigure}[b]{0.49\textwidth}
    \includegraphics[width =\textwidth]{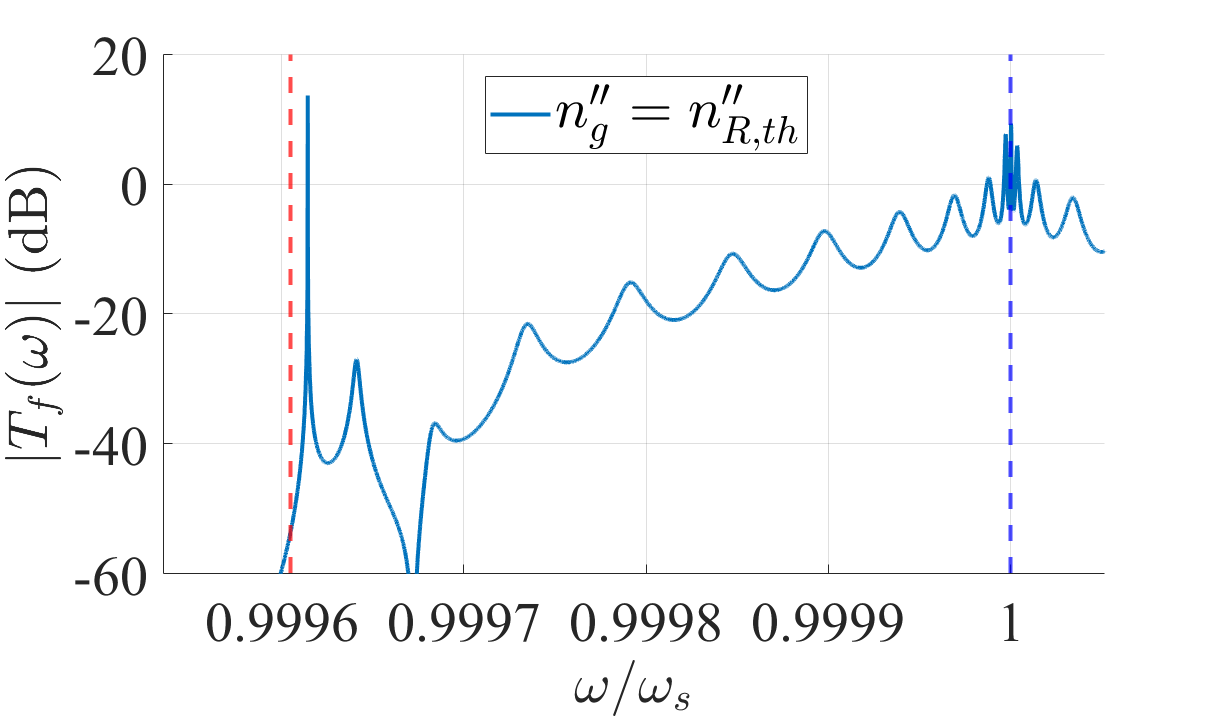}
    \caption{}
  \end{subfigure}
  \caption[\linewidth]{The magnitude of the function $|T_f(\omega)|$ in (dB) for the SIP-ASOW with $N=25$ for different values of gain. The SIP and the RBE frequencies are shown as vertical dashed blue and red lines, respectively. (a) With no gain, resonances accumulate around $\omega_S$. The function $|T_f|$ near the SIP is already higher than near the RBE. (b) With $n_g^{\prime\prime}=n_{S,th}^{\prime\prime}$, the highest peak occurs at the SIP resonance frequency. (c) With $n_g^{\prime\prime}=n_{R,th}^{\prime\prime}$, $|T_f(\omega_{S,res})|$ is reduced and the maximum occurs at the RBE resonance $\omega_{S,res}$.}
  \label{fig:SIP-RBE-TransferThreshold}
\end{figure}

Figure~\ref{fig:SIPRBElasingvsN}(c) shows the lasing threshold against the quality factor of an SIP-ASOW laser operating near the RBE (in black) and near the SIP (in blue), where each dot refers to a finite-length ASOW of $N$ unit cells, with $N\in[5,40]$. Each ASOW has a higher $Q$ at the RBE resonance (nearest resonance to the RBE frequency) than that at the SIP resonance (nearest resonance to the SIP frequency). Therefore, one would a priori expect the RBE lasing threshold to be lower than the SIP lasing threshold. However, the opposite has been observed. This discrepancy may be attributed to the impact of the field amplitude distribution within the underlying passive cavity on the lasing threshold after the inclusion of gain in the cavity. Indeed, Figure~\ref{fig:SIPRBElasingvsN}(c), which compares the SIP-induced lasing threshold with the RBE-induced lasing threshold, shows similar curves than those in Figure~16 in \cite{veysi_degenerate_2018}, which compares a DBE cavity and an RBE cavity with the same quality factors.The authors show that the lasing threshold induced by the DBE is lower than that induced by the RBE because the field amplitude distribution is greater in the DBE cavity. See also a related discussion in \cite{othman_giant_2016} in terms of local density of states distribution within the cavity, for the DBE and RBE cases. However, in this paper, we examine the lasing threshold in the vicinity of two EPDs (an SIP and an RBE) in the same cavity. Our numerical experiments reveal that the SIP-induced lasing threshold is lower than that induced by the RBE, even though the RBE quality factor is larger.Although additional investigation is necessary to understand the difference in the field amplitude distribution in SIP and RBE cavities, and its effect on the lasing threshold, the evidence presented in this paper shows that in an active ASOW, SIP-induced lasing is preferred over lasing near an RBE.

To further illustrate the threshold difference between lasing near the SIP and the RBE, Figure~\ref{fig:SIP-RBE-TransferThreshold} depicts the magnitude of $T_f$ (in dB) against $\omega$ for a finite-length SIP-ASOW with $N=25$ for different values of gain. The vertical blue and red dashed lines depict the SIP and the RBE frequencies, respectively. In Fig.~\ref{fig:SIP-RBE-TransferThreshold}(a) the ASOW has no gain, $n_g^{\prime\prime}=0$. The magnitude of the $T_f$ function at the SIP resonance is already higher than at the RBE resonance in the passive ASOW. Then, one would a priori expect the lasing threshold near the SIP to be lower than near the RBE, as it is indeed the case observed. In Fig.~\ref{fig:SIP-RBE-TransferThreshold}(b) the gain is set to the SIP lasing threshold, $n_g^{\prime\prime}=n_{S,th}^{\prime\prime}$, where the $|T_f|$ experiences a local maximum close to $\omega_S$. The peak at the SIP resonance decreases as the gain increases above its threshold. This effect is explained in detail in Appendix~A. In Fig.~\ref{fig:SIP-RBE-TransferThreshold}(c) the gain is further increased to the RBE lasing threshold, $n_g^{\prime\prime}=n_{R,th}^{\prime\prime}$, and the maximum of $|T_f|$ occurs instead near $\omega_R$. Further investigation is necessary to determine how the enhanced field amplitude distribution in passive cavities operating near an EPD affects $T_f$ and the lasing threshold of the system with gain.
In summary, resonances near the SIP frequency require a smaller effective power gain coefficient $g_{eff}$ to lase than resonances near the RBE frequency. Therefore, though not shown here, we expect that an SIP-based laser has higher lasing efficiency, meaning that it emits more output power with less gain than an RBE-based laser. In all likelihood, the SIP lasing threshold is lower than the RBE lasing threshold due to a combination of factors, which ultimately arise from the contrast between the exceptional properties of the SIP-associated frozen mode and those of the standing wave characteristic of the RBE. Moreover, the threshold difference between lasing near an SIP or an RBE could be increased by using the two degrees of freedom associated with the two left and right terminations, which may affect the RBE and SIP resonances differently.

\section{Conclusion}
\label{ch:Conclusions}

Lasing conditions in the vicinity of an SIP have been investigated for an ASOW terminated on a straight waveguide at each end of the ASOW cavity. The ASOW cavity does not need mirrors to display a high quality factor and low gain threshold. The mode mismatch between the SIP-ASOW and the straight waveguide is attributed to the frozen mode, i.e., to its three degenerate modes and the resulting Bloch impedance. By analyzing the degenerate modes in the periodic ASOW, we have shown that the eigenmode distortion due to the presence of a net gain or loss is more severe near the SIP frequency than at other frequencies; however, the properties associated with the three-mode exceptional degeneracy are in part preserved for relatively small values of gain, which can nevertheless bring the system above threshold. 
We have observed that the quality factor of both the SIP resonance and the RBE resonance scales as a cube of the waveguide length, and the lasing threshold as a negative cube of the waveguide length. While an SIP cavity displays a lower $Q$ factor than an RBE cavity of the same gain medium and length, it also displays a lower lasing threshold. Although we propose that this disparity is caused by the different field distributions in the cavity at the two resonances, it has yet to be confirmed. Nonetheless, our research has revealed that SIP lasing is preferred. While the results here have been shown specifically for the ASOW, the insights and conclusions outlined in this paper can be readily applied to other periodic optical waveguides that display an SIP. Further control of the SIP and RBE lasing thresholds could be exerted by changing the loading at the two ends of the lasing cavity. Future considerations into SIP lasers shall also focus on mode selectivity, the impact of nonlinear effects, and their transient response.

\section*{Acknowledgment}

This research is based upon work supported by the Air Force Office of Scientific Research award numbers LRIR 21RYCOR019 and FA9550-18-1-0355. It has been approved for public release with unlimited distribution.

\section*{Disclosure}
The authors declare no conflicts of interest.

\section*{Data availability statement}
Data underlying the results presented in this paper are not publicly available at this time but may be obtained from the authors upon reasonable request.

\begin{figure}
  \centering
  \begin{subfigure}[t]{0.32\textwidth}
    \includegraphics[width =\textwidth]{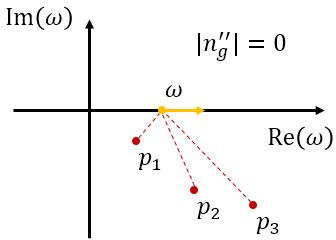}
    \caption{}
  \end{subfigure}
  \begin{subfigure}[t]{0.32\textwidth}
    \includegraphics[width =\textwidth]{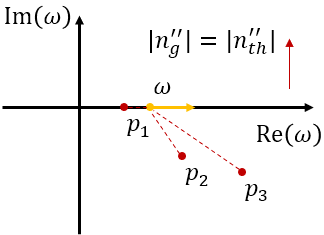}
    \caption{}
  \end{subfigure}
  \begin{subfigure}[t]{0.32\textwidth}
    \includegraphics[width =\textwidth]{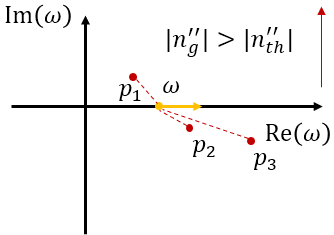}
    \caption{}
  \end{subfigure}
  \caption{Example of three complex poles (red dots) of the transfer function $T_f(\omega)$, mathematically extended also to the case when the system is unstable, i.e., when a pole has $\text{Im}(p_i)>0$. The function $T_f$ is calculated as in Eq. (\ref{eq:TransferFunctionAsPoles}) for real $\omega$ (shown with an orange arrow ) for different values of gain. (a) With no gain, the system is stable. (b) The gain is set at the lasing threshold associated either to $\omega_S$ or $\omega_R$. The transfer function has a maximum because the distance between $\omega$ and the poles is minimum. (c) The gain is increased above the lasing threshold. The value of $T_f(\omega)$ decreases as the distance between the poles and $\omega$ increases.}
  \label{fig:ComplexFreqPlane}
\end{figure}

\section*{Appendix A: Lasing threshold calculations near an SIP and an RBE}

We define the lasing threshold in a finite-length ASOW as the minimum gain required to maintain oscillations in the cavity. It is calculated by gradually increasing $|n_g^{\prime\prime}|$ until the ASOW of finite length terminated on two straight waveguides is marginally stable. Its stability is tracked through the poles $p_i$ of the function $T_f(\omega)$, rewritten as \cite{Franklin_Feedback_2001, nada_exceptional_2020}

\begin{equation}
  T_f (\omega) \propto \frac{1}{\Pi|\omega-p_i|},
  \label{eq:TransferFunctionAsPoles}
\end{equation}

where $\omega$ is the sweeping real frequency. This function is the extension of the transfer function in Eq.~(\ref{eq:Transfer Function}) to the case of complex poles associated with an unstable regime. Since the electric field is a real-valued quantity, poles occur in pairs \cite{nada_exceptional_2020}: if $p_i$ is a pole of the system, $-p_i^*$ is a pole as well. Figure~\ref{fig:ComplexFreqPlane} depicts three poles (the red dots) of the function $T_f(\omega)$ in the complex $\omega$ space for different values of gain. The orange arrow illustrates that $\omega$ sweeps the real axis while we calculate $T_f(\omega)$. 
Figure~\ref{fig:ComplexFreqPlane}(a) shows the stable pole distribution in a system with no gain. 
All the poles of the transfer function in a stable ASOW have negative imaginary parts. The system is unstable when at least one pair of poles has a positive imaginary part. Increasing the gain, which is modeled as $n_g^{\prime\prime}<0$, moves the poles upwards. Operating at a resonance, defined here as the real frequency of the peak, $|T_f(\omega_{res})|$ reaches a maximum when the system is marginally stable \cite{nada_exceptional_2020}, i.e., the lasing threshold is the gain that renders the first pair of poles such that $\text{Im}(p_i)=0$. Figure~\ref{fig:ComplexFreqPlane}(b) depicts the complex $\omega$ space for a marginally stable waveguide, where the gain is set at the lasing threshold. Therefore, there is a real $\omega$ such that $|\omega-p_i|=0$, which renders $|T_f|\rightarrow \infty$. If this happens near the SIP frequency we say we have an SIP laser, if this peak happens near the RBE frequency we say we have an RBE laser. 
Further adding gain moves those poles into the unstable region, reducing $|T_f|$. Figure~\ref{fig:ComplexFreqPlane}(c) depicts a pole distribution for gain above the lasing threshold (depicted by a larger vertical red arrow on the right of the figure). The resonator made of the finite-length ASOW terminated onto two straight waveguides is unstable and $|T_f(\omega)|$ would be reduced as the distance between any $\omega$ on the real axis and the closest pole has increased. Each plot in Fig.~\ref{fig:SIP-RBE-TransferThreshold} corresponds with a stability region in the three scenarios depicted in Fig.~\ref{fig:ComplexFreqPlane} when considering the SIP resonance. Fig.~\ref{fig:SIP-RBE-TransferThreshold} also shows the peak associated with the threshold at the RBE resonance, happening for a gain value larger than the SIP threshold. 

Losses increase the lasing threshold because they move the poles downwards.

\bibliography{references}

\end{document}